\documentclass[]{iopart}
\usepackage{iopams}
\usepackage{epsfig,amssymb, amsthm,  bm} 

\newcommand{\be}{\begin{eqnarray}}
\newcommand{\ee}{\end{eqnarray}}

\newcommand{\finpr}{\hfill $\square$ \vspace{2mm}}

\def\be{\begin{eqnarray}}
\def\ee{\end{eqnarray}}
\def\bee{\begin{eqnarray*}}
\def\eee{\end{eqnarray*}}
\newtheorem{thm}{Theorem}
\newtheorem{cor}{Corollary}

\newtheorem{obs}{Observation}

\newtheorem{defn}{Definition}

          \def\tr{\hbox{Tr}}

\theoremstyle{definition}\newtheorem{rem}{Remark}

\begin{document}



\title{Fundamentals of universality in one-way quantum computation}
\author{M. Van den Nest$^1$, W. D\"ur$^{1,2}$, A. Miyake$^{1,2}$\\ and H. J. Briegel$^{1,2}$}

\address{$^1$ Institut f\"ur Quantenoptik und Quanteninformation der \"Osterreichischen Akademie der Wissenschaften, Innsbruck, Austria\\
$^2$ Institut f{\"u}r Theoretische Physik, Universit{\"a}t Innsbruck,
Technikerstra{\ss}e 25, A-6020 Innsbruck, Austria
}
\date{\today}

\begin{abstract}

In this article, we build a framework allowing for a
systematic investigation of the fundamental issue:
\emph{``Which quantum states serve as universal resources
for measurement-based (one-way) quantum computation?''} We
start our study by re-examining what is exactly meant by
``universality'' in quantum computation, and what the
implications are for universal one-way quantum computation.
Given the framework of a measurement-based quantum
computer, where quantum information is processed by
\emph{local} operations only, we find that the most general
universal one-way quantum computer is one which is capable
of accepting arbitrary classical inputs and producing
arbitrary quantum outputs---we refer to this property as
\emph{CQ-universality}. We then show that a systematic
study of CQ-universality in one-way quantum computation is
possible by identifying entanglement features that are
required to be present in every universal resource. In
particular, we find that a large class of entanglement
measures must reach its supremum on every universal
resource. These insights are used to identify several
families of states as being not universal, such as 1D
cluster states, GHZ states, W states, and ground states of
non-critical 1D spin systems. Our criteria are strengthened
by considering the efficiency of a quantum computation, and
we find that entanglement measures must obey a certain
scaling law with the system size for all efficient
universal resources. This again leads to examples of
non-universal resources, such as, e.g., ground states of
critical 1D spin systems. On the other hand, we provide
several examples of efficient universal resources, namely
graph states corresponding to hexagonal, triangular and
Kagome lattices. Finally, we consider the more general
notion of encoded CQ-universality, where quantum outputs
are allowed to be produced in an encoded form. Again we
provide entanglement-based criteria for encoded
universality. Moreover, we present a general procedure to
construct encoded universal resources.

\end{abstract}

\pacs{03.67.Lx, 03.67.Mn, 03.65.Ta}

\maketitle

\tableofcontents


\section{Introduction}

\subsection{Role of MQC in fundamental investigations}

Quantum computers offer a promising new way of information
processing, in which the distinguished features of quantum
mechanics can  fruitfully be exploited. The discovery of
quantum algorithms, most notably Shor's factoring algorithm
\cite{Sh94} and Grover's search algorithm \cite{Gr97},
demonstrate that quantum computation can achieve a
(possibly exponential) speed-up over classical devices.
This has put quantum computation at the focus of
contemporary research, and, indeed, significant progress in
the theoretical understanding of quantum information
processing, as well as promising steps toward an
experimental realization of large-scale quantum
computation, have recently been reported. In spite of these
exciting developments, the basic question: '{\em Which
features of quantum mechanics are responsible for the
speed-up of quantum computation over classical devices?}'
remains to date largely unanswered.

An indication that there might not be a straightforward
answer to this fundamental but difficult question, is given
by the fact that there exist various models for quantum
computation, such the quantum Turing machine \cite{De85},
the quantum circuit (or network) model \cite{De89},
measurement-based models \cite{Ra01, Ra03, Go99, Ni03,
Pe04, Da04, Ve04, Gr06}, as well as adiabatic quantum
computation \cite{Fa01}. Although these models have been
shown to be equivalent in a certain complexity-theoretic
sense (colloquially speaking, any problem which can be
solved efficiently in one of the above models, can also be
solved efficiently by the others), the elementary concepts
underlying these schemes may differ significantly.

Therefore, certain computational schemes may lend
themselves more than others  to understand fundamental
issues regarding the power of quantum as compared to
classical computation.  The new paradigm of {\em
measurement-based quantum computation (MQC)}, with the
one-way quantum computer \cite{Ra01, Ra03} and the
teleportation-based model \cite{Go99, Ni03} as the most
prominent examples, has lead to fresh perspectives in these
respects. While in, e.g., the circuit model quantum
information is processed by coherent unitary evolutions, in
MQC the processing of quantum information takes place by
performing \emph{sequences of adaptive measurements}.
Teleportation-based models use joint (i.e., entangling)
measurements on two or more qubits and thereby perform
sequences of teleportation-based gates. In contrast, the
one-way model  uses a highly entangled state, the
\emph{cluster state} \cite{Br01}, as a universal resource
which is processed by single-qubit measurements only.

The latter property provides the model of one-way quantum
computation---which is the focus of this article---with a
very distinct feature, namely that the entire resource of
the computation is provided by the entangled state in which
the system is initialized. In particular, any computational
speed-up of such a model w.r.t. classical computation can
be traced back entirely to the properties of the resource
state.

What is more, in the one-way model the resource character
of \emph{entanglement} is particularly highlighted, as it
is clearly separated from the processing of quantum
information by local, single-qubit measurements. As the
latter cannot increase any entanglement in the system, all
entanglement required for quantum computation needs to be
initially present in the system.  The introductory question
of this article can therefore be rephrased in a more
concise form, namely {\em 'What are the essential
entanglement features of a resource state for (one-way) MQC that are
required to obtain a speed-up over classical computation?'}
The present article will be centered around this question.

We emphasize that, in the following, we will exclusively
consider MQC in the sense of the one-way quantum computer,
i.e., considering only local measurements;
teleportation-based models will not be considered.

For proposed implementations of measurement-based
computation see e.g. Refs. \cite{Ra03, Br02,  Bo05, Cl05,
HybridQC, Ta06a, LOQC, Me06, Ki06} (further
references can be found in Ref.~\cite{He06}); for recent
experimental developments see Refs. \cite{exp}.

\subsection{Universality in MQC---aim and contribution of this paper}

In this paper, we will be interested in those resource
states for MQC which possess the strongest computational
power possible, namely those which enable \emph{universal}
quantum computation, as do the  2D cluster states
\footnote{For completeness we mention that we will
\emph{not} investigate whether universal
(measurement-based) quantum computation can be simulated
efficiently on a classical computer, which remains to date
an unresolved issue. In other words, here we study
universal resource states---having the maximal
computational power which a quantum computer can have---but
we do not investigate whether this maximal computational
power is in fact stronger than classical computation. For
studies of the possibility of classically simulating
quantum computation, we refer to Refs. \cite{sim, Va06a}.}.
It is our aim to gain insight under which conditions
resource states, other than the 2D cluster states, are
universal, and what the role of entanglement plays in this
issue.

In order to study universality in MQC in detail, it is
necessary to have a clearcut definition of this notion. We
will hence start our study (Sec. \ref{General}) by
re-examining what is exactly meant by ``universal quantum
computation'', and what the implications are for
universality in MQC (Sec. \ref{sect_universal_resource}).
We will find that the notion of universality strongly
depends on whether the in- and outputs  of a quantum
computer are allowed to be either classical or quantum.
Furthermore, we will argue that the most general universal
one-way quantum computer is a device which accepts
arbitrary classical (C) inputs and produces arbitrary
quantum (Q) outputs---we will call such a device a
\emph{CQ-universal} one-way quantum computer. The property
that a one-way quantum computer is restricted to accept
classical information only, is essentially due to the fact
that, after the resource state of the device has been
prepared, the only allowed quantum operations are
\emph{local operations}, which do not allow one to couple in quantum
input states. One the other hand, the generation of
arbitrary quantum output states poses no problem (as is the
case in the 2D cluster state model)---hence the notion of
CQ-universality in MQC. As a standard example, the 2D
cluster state model is CQ-universal.

After giving a formal definition of what constitutes a
\mbox{(CQ-)}universal resource state for MQC, we develop a
framework which allows a systematic investigation of the
criteria which need to be met by every universal resource
(Sec. \ref{sect_ED_criteria}). Our approach, which has been
initiated in Ref. \cite{Va06}, is centered around
considerations regarding entanglement. In particular, we
find that every universal resource needs to be
\emph{maximally entangled}, in the sense that every
entanglement measure (belonging to a well-defined class)
must reach its supremum on every universal resource. This
result subsequently leads to several criteria for
universality, by applying the result to specific measures.
We consider several examples and show that entropic
entanglement width \cite{Va06}, Schmidt rank width \cite{Va06a}, Schmidt measure \cite{Ei01},
geometric measure of entanglement \cite{Sh95, We03} as well as measures
describing the capability to generate Bell pairs, must grow
unboundedly on every universal resource---thus, every
resource which does not exhibit a divergence of the above
measures, cannot be universal. This leads to several
examples of states which are not universal, such as large
families of graph states, the GHZ states, W states and
ground states of non-critical strongly interacting 1D
systems.

Along the way, we also take {\em efficiency} into account
(Sec. \ref{sect_I_eff}). In this case, we find that a
necessary condition for efficient universality is that the
growth of above mentioned entanglement measures must be
sufficiently fast with the system size---in most cases
faster-than-logarithmic. These results are again used to
provide further examples of states which are not
efficiently universal. Furthermore, we also construct
examples of efficient universal resource states, namely
graph states corresponding to 2D hexagonal, triangular and
Kagome lattices (Sec. \ref{go}). Here our proof consists of
showing explicitly that all these resources can be
transformed into each other (with a certain moderate
overhead) by local operations and classical communication
(LOCC)---and are therefore equally maximally entangled.

In Sec. \ref{encoded}, we consider a weaker (i.e., more
general) form of CQ-universality, namely \emph{encoded
universality}, where it is sufficient that the desired
output states be generated in an {\em encoded} form (here
the notion of ``encoding'' is similar to e.g. schemes in
fault-tolerance and quantum error-correction). After
discussing encodings and providing a definition for encoded
universality, we investigate to which extent
entanglement-based criteria can be used to assess encoded
non-universality. Most importantly, we show that the
Schmidt measure and geometric measure, as well as all
measures which are non-increasing under coarsening of
partitions of the system, give rise to criteria for encoded
universality. Furthermore, relying on an ``indirect''
argument of classical simulatability of quantum
computation, we also argue that the Schmidt rank width
provides a criterion for encoded universality. We then also
give general  constructive results; most importantly,
extending results put forward in Ref. \cite{Gr06}, we find
that any universal resource which is itself subsequently
encoded in an arbitrary way, is an encoded universal
resource---which is a nontrivial statement. This implies in
particular that for {\em any} kind of encoding, an encoded
2D-cluster state is an encoded universal resource (up to
logical Pauli operations). We provide extensions to encoded
universality and discuss some examples of encoded universal
resources presented in the literature.

Finally, a summary of our results and an outlook toward
further investigations are formulated in Sec.
\ref{sect_summary} and \ref{sect_outlook}.

\begin{rem}The main text is supplemented with a number of remarks. Although these remarks provide
information which is relevant to obtain a detailed
understanding of the current investigation, they may
be skipped in a first reading. \hfill $\diamond$
\end{rem}

\section{General considerations on universality}\label{General}

In this section, we discuss what a universal quantum
computer should be capable of. Somewhat unexpectedly, this
is not so trivial as one might think at first sight. In
particular, we will argue that there are several possible
definitions, all of which are of natural interest depending
on which application one might have in mind.

We will start our discussion by reviewing, in section \ref{sect_network},
some important conceptual notions regarding the
universality of the circuit model. In investigations of
universality for general computational models, the circuit model will serve as a natural
reference. This then leads us, in section \ref{sect_types_uni}, to consider
four natural but distinct notions of universality of a
quantum computer, namely {\em CC-, CQ-, QC}, and
{\em QQ-universality}. In each of those notions of universality
it is specified whether the computer should accept
quantum (Q) or classical (C) inputs, and produce quantum
or classical outputs.

\subsection{Circuit model}\label{sect_network}

In the circuit model for quantum  computation, quantum
information is processed by applying sequences of unitary
gates on quantum states. These gates are typically chosen
from an elementary set. Universality in the circuit model
is defined with respect to the possibility of generating
arbitrary unitary operations by composition of such
elementary gates. Namely, a gate set ${\cal S}$ is called
{\em universal} if {\em any} $n$-qubit unitary operation
$U\in SU(2^n)$, for every $n$, can be realized as a
sequence of elementary gates taken from ${\cal S}$.

Notice that in principle the number of gates required to
produce a given unitary may be unbounded. In practice,
however, the number of gates that can be applied is
limited, and one often considers only $n$-qubit unitary
operations $U$ that can be generated \emph{efficiently},
i.e., with poly$(n)$ elementary gates. An example of a
universal gate set ${\cal S}_1$ is given by the two-qubit
CNOT gate together with the group $SU(2)$ of arbitrary
single-qubit gates \cite{NiCh98}, and this specific gate
set usually serves as a reference set to distinguish
between efficient and non-efficient gate sets. In
particular, one says that a set of gates ${\cal S}$ is
efficiently universal if any $U \in SU(2^n)$ that can be
efficiently generated---i.e., with poly($n$) gates---using
the gate set ${\cal S}_1$, can also be efficiently
generated using the gate set ${\cal S}$. This implies that
efficient universal gate sets can simulate the set ${\cal
S}_1$ with polynomial overhead.

For many practical applications, the exact generation of a
unitary operator is not required, and an {\em approximate}
application of $U$ with a certain accuracy $\epsilon$ is
sufficient. Such a finite accuracy shows up very naturally
when considering discrete sets of elementary gates, such as
the CNOT gate plus a finite set of single-qubit gates with
certain rotational angles \cite{NiCh98}. In fact, it turns out that an
approximate generation of all unitary operations with
arbitrary accuracy is possible for many discrete sets of
elementary gates. The best known examples is the set
consisting of single- and two-qubit Clifford gates
\cite{Go97} together with a single-qubit $\pi/8$ rotation
along the $z$-axis \cite{NiCh98}.
As established in the Solovay-Kitaev theorem, the
overhead to approximate a sequence of $m$ elementary gates
from the set ${\cal S}_1$ with accuracy $\epsilon$ by gates
from the discrete gate set specified above, scales as $O(m
\log^c(m/\epsilon))$, i.e., poly-logarithmically in
$1/\epsilon$ (where $1\leq c\leq 2$) \cite{NiCh98, Da05}.

It is evident from above discussion that universality in
the circuit model is concerned with the possibility of
implementing arbitrary {\em unitary operations}. However,
regarding the nature of a quantum computation in the
circuit model, often additional (implicit) assumptions are
made concerning both the {\em input} and {\em output} of a
computation. Namely, it is often assumed that both the
input and the output of a quantum computation are
classical. This manifests itself in the fact that, first,
the input state of a computation is typically a standard
quantum state $|0\rangle^{\otimes n}$ and, second, the last
step of the computation---after the unitary operation has
been implemented---typically consists of a sequence of
single--qubit measurements in the standard basis,
destroying the final quantum state (i.e., transforming it
to a simple product state) and yielding the classical
output. We emphasize that it is {\em not} necessary to
restrict a quantum computation in the circuit model to this
classical input--output scenario. In particular, arbitrary
quantum inputs in the form of $n$-qubit quantum states
$|\psi\rangle$ can naturally be processed by a (universal)
circuit model quantum computer. In addition, it is  not
necessary to perform a final measurement. Without such
final measurements, the quantum computation produces a
quantum state as output, which might be used for other
purposes. In this general scenario, a circuit model quantum
computer might be considered as a device which accepts a
quantum state as an input, processes it by applying a
unitary operation (representing the program and possible
classical input data), and finally produces  an output
quantum state (together with classical information
resulting from possible measurements).

Thus, a (universal) circuit model quantum computer can be
regarded as a device which accepts either a classical or
quantum input, and which produces either a classical or
quantum output. As we will discuss in more detail in
section \ref{sect_types_uni}, the different choices one can
make regarding the types of in- and outputs of a quantum
computation have a significant impact on the definition(s)
of universality one may consider. This will be a crucial
point in defining and studying universality in
measurement-based quantum computation.

\subsection{Different types of universality}\label{sect_types_uni}

As we have briefly reviewed in the previous section,
universality in the circuit model is concerned with the
possibility of implementing arbitrary {\em unitary
operations}---be it exactly or with a certain precision.

When considering a quantum computational model other than
the circuit model---such as the measurement-based model
considered here---one can ask in which sense it may be
called universal. In order to define universality, one
typically takes the circuit model as a reference, and this
is also the strategy which will be adopted here
\footnote{This strategy is natural because all ``standard''
models for QC are known to be equivalent, in a
complexity-theoretic sense, to the circuit model.}.
Colloquially speaking, a model is called universal {\em if
it can perform the same tasks as a circuit model quantum
computer}. In this section, we make this statement more
precise (although we will stay at a qualitative level). In
particular, we will argue that possible definitions of what
a universal quantum computer is supposed to be capable of,
depend on the allowed input and output, which may either be
of classical or quantum nature.

Regarding in- and outputs of a quantum computation, we have the following possibilities:
\begin{itemize}
\item[{\bf  CC}:] Both input and output are classical;
\item[{\bf QC}:] The input is quantum and the output is
classical;
\item[{\bf CQ}:] The input is classical and the output is a quantum;  
\item[{\bf QQ}:] Both input and output are quantum.
\end{itemize}
The notion of ``universal quantum computation'' has different
meanings depending on which of the above cases is
considered. In each of these situations, a natural
definition of universality can be formulated where the
circuit model is used as a reference. Moreover, we
emphasize that we will make a clear separation between
universality of a quantum computer and the efficiency with
which computations can be executed---i.e., we will regard
\emph{universality} and {\em efficient universality} as
distinct notions (this point will be discussed further in
section \ref{sect_remarks_uni}).

We now consider the cases CC, CQ, QC and QQ one by one.

{\bf CC.---} Here it is assumed that both input and output
of the computation are {\em classical}, i.e., a quantum
computer is considered to be a device that solves classical
problems and provides the solution in the form of classical
data. Only at an intermediate stage the additional power of
quantum mechanics is used to achieve an enhanced processing
of information. The final step of such a quantum
computation is usually given by a sequence of (local)
measurements performed on the output state, which ensures
the transition from quantum information to classical
information. An important example of this kind  is given by
Shor's factoring algorithm.
One may call a quantum computer CC-universal
if it can perform the same tasks as a circuit model quantum
computer which accepts classical inputs and produces
classical outputs. That is, for every unitary $U$, a
CC-universal quantum computer must be capable of
reproducing the statistics of local measurements performed
on the state $U|0\rangle^{\otimes n}$. Note that, as we do
not take efficiency into account yet, a CC-universal
quantum computer is simply one which can perform universal
classical computation (and is therefore not different from
a classical computer). In order to distinguish a
CC-universal quantum computer from classical computers, one
needs to take efficiency into account and demand that the
classical data is obtained in polynomial time for any
unitary operations $U$ that can be generated with a
polynomial sized quantum circuit. We refer to remark
\ref{remark_eff} for further details regarding this issue.

{\bf QC.---} In this case, the output is still classical,
i.e., the final step of the computation is given by a
sequence of measurements. However, now the input can be an
arbitrary quantum state $|\phi_{\rm in}\rangle$ that is
processed by the quantum computer. The input state may be
the output state of a previous QQ-quantum computer (see the
case QQ below), or the state of some physical system which
one would like to study with the help of a (QC) quantum
computer, e.g., the ground state of some spin system. For
instance, one may wish to learn the value of some
observable or entanglement measure, or the fact whether a
state (or density operator) has positive partial
transposition \cite{Pe93}. Note that the input state may be
known or unknown to the device. Finally, also in studies of
quantum complexity classes such as QMA, situations are
considered where devices can accept quantum states as
inputs ("certificates") \cite{Ah02}.

A quantum computer is then called QC-universal if it can
perform the same tasks as a circuit model quantum computer
which accepts quantum inputs and produces classical
outputs. That is, {\em for every (possibly unknown) input
state $|\phi_{\rm in}\rangle$ and for every unitary
operation $U$, a QC-universal quantum computer can
reproduce the statistics of local measurements performed on
the quantum state $|\phi_{\rm out}\rangle = U |\phi_{\rm
in}\rangle$}. One may in addition take efficiency into
account, in the same way as in the case of CC, demanding
that the overhead with respect to the circuit model is
polynomial \footnote{However, for distinguishing such a
device from classical computers this is not necessary (a
classical computer cannot handle a quantum input).}.

{\bf CQ.---} Here the input is classical but the output of
a quantum computation can now be a quantum state. Compared
to a CC quantum computer, this device has the advantage
that one can decide at a later stage whether one wishes to
obtain classical information by performing measurements, or
whether one wants to use the produced state as the input of
a following quantum computation, or whether one wants to
use this state for another quantum information task---the
goal of a quantum computation might e.g. be to produce a
quantum state which is subsequently distributed over
several parties, and then used to establish a secret key,
for secret sharing, or to perform (non-local) two-qubit
gates.  

Analogous to the previous cases, we will call a quantum
computer  CQ-universal if it can perform the same tasks as
a circuit model quantum computer which accepts classical
inputs and which produces quantum outputs. That is, {\em
for every unitary operation $U$, a CQ-universal quantum
computer is able to produce the quantum state $|\phi_{\rm
out}\rangle  = U |0\rangle^{\otimes n}$}. This definition
is equivalent to stating that that a CQ-universal quantum
computer is capable of preparing an arbitrary quantum
state.
One may again add the requirement of efficiency here,
calling a quantum computer efficiently CQ-universal if the
overhead in the above state preparation compared to the
circuit model is polynomial.

{\bf QQ.---} This is the most powerful device, as it can
accept quantum states as input and can produce quantum
states as output. Apart from the applications mentioned in
(ii) and (iii), such a quantum computer is {\em
composable}, i.e., the output of a previous quantum
computation can be used as the input of a subsequent
quantum computation. In particular this allows one to use
distributed quantum processors. We will call a quantum
computer QQ-universal if  it can produce the quantum state
$|\phi_{\rm out}\rangle  = U |\phi_{\rm in}\rangle$ for any
given input state $|\phi_{\rm in}\rangle$ and for any
unitary operation $U$. Equivalently, one can say that any
unitary operation $U$ can be performed on an arbitrary
input state. As before, the notion of efficiency can
naturally be considered.

\subsection{Some remarks}\label{sect_remarks_uni}

In this section, we give a number of remarks regarding the
above types of universality.

\begin{rem} {\it Separating universality and efficiency.---}\label{remark_eff} In the
definitions of CC-, \mbox{CQ-}, QC- and QQ-universality we
have deliberately  separated the issue of universality from
the issue of efficiency. The reason for this is that, in
the present study of MQC, we will be interested in
computational models having a quantum output. In such
situations, we will argue that a lot can be learned by
studying, say, CQ-universality, without considering
efficiency. This will be in particular the case for the
study of necessary conditions for CQ-universality: we will
present systematic procedures for identifying large classes
of resource states for MQC which are not CQ-universal, even
though no considerations of efficiency are included.
Finally, we note that in the case of CC-universality it is
meaningless to separate universality from efficiency; as
pointed out above, every classical computer is
(inefficiently) CC-universal, such that nothing can be
learned about the differences between classical and quantum
computation by considering CC-universality without
efficiency.  \hfill $\diamond$ \end{rem}

\begin{rem} {\it Qubits, qudits and encodings.---} In the
above definitions of the different types of universalities,
and in the discussion of the circuit model, we have always
assumed that the physical systems involved are qubit
systems (and qubit systems only). This does not represent a
restriction if one considers classical inputs and outputs,
as only the information is important in this case; the type
and dimension of systems that carry the information is
irrelevant \footnote{We note that, in the CC case, the
circuit model working with qudit systems has been shown to
polynomially equivalent to the qubit circuit model
\cite{NiCh98}.}. The situation is  different when
considering also quantum inputs and/or outputs---i.e., in
the CQ, QC, and QQ case. There, any computational model
typically works with physical systems where the individual
constituents are associated to a  Hilbert space  of fixed
dimension, such that the states which can be accepted
and/or produced can only be multi-party $d$-dimensional
systems, for some fixed $d$. For example, it is clear that,
when local measurements are performed on a cluster state in
the one-way model, only qubit states can be produced, and
not e.g. qutrit states.

Notice, however, that in many cases it is sufficient to generate or provide quantum states in an {\em encoded} form. That is, the Hilbert space of a $d$-dimensional system can be embedded into the Hilbert space of $\lceil\log_2(d)\rceil$ qubits and in this sense any state of $n$ $d$-dimensional systems can be generated in an encoded form with $n'= n \cdot\lceil\log_2(d)\rceil$ qubits. We will discuss the issue of encodings in more detail in Sec. \ref{encoded}. Until then we will restrict our considerations to qubits, keeping in mind that higher dimensional systems can be produced in an encoded form.
\hfill $\diamond$ \end{rem}

\begin{rem}\label{remark_unknown_input} {\it Known versus unknown quantum input.---}
A QC- or QQ-universal quantum computer must be capable of
processing unknown inputs. Furthermore,  although a
CQ-universal computer can only accept classical inputs, it
can simulate processes where a QC or QQ computer accepts a
{\em known} quantum input: namely, suppose that a
QQ-universal computer accepts a known input state
$|\phi_{\rm in}\rangle$ and performs a unitary operation
$U$,
producing the state $|\phi_{\rm out}\rangle  = U |\phi_{\rm in}\rangle$. Then a CQ-universal computer can simulate this process by simply classically computing what the state $|\phi_{\rm out}\rangle$ is, and by then preparing this state, which is possible from its CQ-universality. When efficiency is taken into account, an efficient CQ-universal computer can efficiently simulate any process $|\phi_{\rm in}\rangle  \to |\phi_{\rm out}\rangle$ where (i) $U$ is a poly-sized circuit, (ii) $|\phi_{\rm in}\rangle$ can be prepared by a poly-sized circuit and (iii) this last circuit is known; for, in this case $|\phi_{\rm out}\rangle$ is simply a state which can be prepared by a known poly-sized circuit, which any efficiently CQ-universal resource should be able to generate efficiently.  
\hfill $\diamond$\end{rem}

\begin{rem} {\it Exact versus approximate
universality.---} In the qualitative treatment in section
\ref{sect_types_uni}, we have considered an ideal situation
and not talked about approximate universality. This is of
course an important issue---cf. e.g. the circuit model,
where finite sets of elementary gates can only be
approximately universal. Nevertheless, when considering MQC
models, it is known that e.g. the 2D cluster states are
exactly \mbox{(CQ-)}universal, and we will consider only
this ideal situation in this paper. Approximate
universality will be taken into account in upcoming work
\cite{Mo07}. \hfill $\diamond$ \end{rem}

\section{Universality in MQC}\label{sect_universal_resource}

In this section, we move to our topic of interest, namely
(one-way) measurement-based models of quantum computation.
In section \ref{sect_uni_oneway}, we will consider the
one-way model with a 2D cluster state as a resource state,
and we will review in which ways it is universal. In fact,
we will see that the 2D cluster state model is
(efficiently) CQ-universal, and that this is the most
powerful type of universality any  MQC model can have (if
we only consider local measurements and do not allow for
additional resources). This leads us to introduce, in
section \ref{sect_defn_uni}, the definition of ``universal
resource for measurement-based quantum computation'', which
refers to CQ-universality. Afterward we discuss the notion
of efficiency in universal MQC.

\subsection{One-way model}\label{sect_uni_oneway}

In this section, we first briefly discuss the distinct
features of the one-way model for MQC having a 2D cluster
state as an entangled resource, and then consider in which
way(s) it is universal.

The one-way quantum computer  was introduced in Ref.
\cite{Ra01, Ra03}. This computational model is in striking
contrast to the circuit model, as the processing of quantum
information is not realized by applying unitary gates
(i.e., via physical interactions between the qubits), but
it takes place solely by performing {\em single-qubit
measurements}. The general procedure of a one-way quantum
computation is the following:
\begin{itemize}
\item[(i)] a classical input is provided which specifies the data and the program;
 \item[(ii)] A 2D cluster state $|C_{d_1\times
d_2}\rangle$ of size $d_1 \times d_2$ is prepared, where
$d_1$ and $d_2$ depend on the classical input data, the
size of output and the length of the computation. A cluster
state is a particular instance of a \emph{graph state}. A
graph state on $m$ qubits is the joint eigenstate of $m$
commuting correlation operators \be K_a:=
\sigma_x^{(a)}\bigotimes_{b\in N(a)} \sigma_z^{(b)},\ee
where $N(a)$ denotes the set of neighbors of qubit $a$ in
the graph \cite{He06}; a 2D cluster state is obtained if
the underlying graph is a $d_1\times d_2$ rectangular
lattice (thus $m=d_1d_2$).  The cluster state serves as the
resource for the computation. \item[(iii)]  A sequence of
adaptive one-qubit measurements is implemented on certain
subsets of qubits in the cluster. In each step of the
computation, the measurement bases depend on the program
and on outcomes of previous measurements. A simple
classical computer is used to compute which measurement
directions have to be chosen in every step of the
computation. \item[(iv)] After the measurements have been
implemented, the state of the system has the form
$|\xi^{\alpha}\rangle |\psi^{\alpha}_{\mbox{\scriptsize
out}}\rangle$, where $\alpha$ indexes the collection of
measurement outcomes of the different  branches of the
computation. The states $|\psi^{\alpha}_{\mbox{\scriptsize
out}}\rangle$ in all branches are equal up to a (local)
discrete Pauli unitary operation, i.e., there exists a state
$|\phi_{\mbox{\scriptsize out}}\rangle$ such that $
|\psi^{\alpha}_{\mbox{\scriptsize
out}}\rangle=\Sigma^{\alpha}|\phi_{\mbox{\scriptsize
out}}\rangle$ for all $\alpha$, where $\Sigma^{\alpha}$ is
a multi-qubit Pauli operator, the so--called byproduct
operator; the measured qubits are in a product state
$|\xi^{\alpha}\rangle$ which also depends on the
measurement outcomes.
\end{itemize}

Thus, in the one-way model every desired state
$|\phi_{\mbox{\scriptsize out}}\rangle$ can be prepared
deterministically up to a Pauli operator---even though the
results of the measurements are random. Only the correction
operations (i.e., the local Pauli operations) depend on the
measurement outcomes, and can be determined via
side-processing with a classical computer.

Let us now discuss the different types of universality for the one-way model.

{\bf QC, QQ.---} First, note that the  one-way quantum
computer in its above form cannot be QC- or QQ-universal by definition. This
is because, in this scheme, one is given a 2D cluster state as a resource and the only allowed quantum
operations are single-qubit measurements corresponding to the classical program. Such local operations do not
allow the device to accept and process a quantum state.

{\bf CC.---} On the other hand, the  one-way model is known
to be {\em efficiently CC-universal}, as was it was proven
in Ref. \cite{Ra01, Ra03} that a one-way quantum computer
can efficiently simulate any CC-computation in the circuit
model. By ``efficiently'' is meant here that the number of
measurements as well as the number of additional qubits and
the temporal overhead required to simulate the
transformation $|0\rangle^{\otimes n}\to
U|0\rangle^{\otimes n}$ scale polynomially with the number
of gates required to generate the unitary $U$ (we refer the
reader to section \ref{sect_def_eff} for a deeper treatment
of efficiency in MQC). Note also that the fact that, in the
one-way model, output states can be prepared up to local
Pauli operators does not cause any restriction, as these
Pauli corrections can be incorporated by accordingly
changing the final measurement bases in the algorithm.

{\bf CQ.---} Even stronger,  with a slight modification the
one-way model can be made to be {\em CQ-universal} and even
{\em efficiently CQ-universal}. This is achieved by
allowing as basic operations, next to the local
measurements, also (local) Pauli operations.  If such local
unitaries are allowed, then it follows from the above that
any multi-qubit state can deterministically be prepared by
performing local measurements on a sufficiently large 2D
cluster state, making the model CQ-universal. Efficient
CQ-universality is obtained by again noting that the
transformation $|0\rangle^{\otimes n}\to
U|0\rangle^{\otimes n}$ can be simulated with polynomial
overhead w.r.t. the circuit model.

\begin{rem} {\it Making the one-way model
QQ-universal.---} As pointed out in remark
\ref{remark_unknown_input}, it follows from the efficient
CQ-universality of the one-way model that it can
efficiently simulate certain processes where a known input
state is transformed by  poly-sized unitary operation. Here
we further remark that the one-way model can be made fully
QQ-universal by allowing an additional resource, an ``input
coupler``, which allows one to couple in unknown quantum
states \footnote{See also ``Scheme 1'' ($\cong$ QQ) versus
``Scheme 2'' ($\cong$ CQ) in Ref. \cite{Ra03}.}. When
$|\phi_{\rm in}\rangle$ is a (possibly unknown) input state
on $n$ qubits, this additional resource simply consists of
$n$ controlled phase gates $U_{\rm PG}=diag(1,1,1,-1)$
which are applied---initially, and only once---pairwise
between every qubit of $|\phi_{\rm in}\rangle$  and every
second qubit on the ``left side'' of a $2n\times m$ cluster
state (for some $m=$ poly$(n)$). It was shown in Ref.
\cite{Ra01, Ra03} that for every unitary operation $U$
there exists a sequence of local measurements, implemented
on the resulting state, which can generate the state
$U|\phi_{\rm in}\rangle$ on some subset of the cluster with
polynomial overhead---making the model efficiently
QQ-universal. Note that a particularly nice feature of this
result is that the measurement protocol is independent of
the input state and only depends on the unitary operation
$U$. Finally, we emphasize that this scheme goes somewhat
beyond the present investigation, as the controlled phase
gates introduce additional entanglement in the system,
going away from the LOCC-only scenario in which we are
interested here (see, however, observation \ref{obs2} in
section \ref{subsect_obs2}) \hfill $\diamond$ \end{rem}

\subsection{General definition}\label{sect_defn_uni}

We conclude from the above discussion that the most general
universality exhibited by the one-way model is (efficient)
CQ-universality. This is the situation we will have in mind
when studying general universality in MQC---i.e., we will
be interested in CQ-universality of general resource
states. The main motivation for this is that
CQ-universality is the most general and powerful type of
universality any LOCC-based  MQC model can have. In particular,
any resource which is (efficiently) CQ-universal is also
(efficiently) CC-universal, and, as we shall see in
observation \ref{obs2}, it is efficiently QQ universal if
supplemented with an input coupler. Furthermore, we will
show in section \ref{sect_ED_criteria} that a systematic
study of the properties which make a resource CQ-universal,
is possible, whereas systematic treatments of e.g.
CC-universality in MQC are presently far less within reach.
{\em Henceforth, when referring to ``universality'' we will
always mean ``CQ-universality''.}

We now have a computational model in mind where quantum
algorithms  are implemented by performing local
measurements (or, more generally, LOCC---see remark \ref{rem_LOCC}) on a resource state, as in the cluster state
model, and we will give a qualitative definition of a
universal resource.

If one wants to avoid talking about infinitely large
states, universality is to be regarded as a property that
is attributed, not to a single state, but to a family of
infinitely many states \be\Psi =\{|\psi_1\rangle,
|\psi_2\rangle, \dots\}, \quad |\Psi|=\infty.\ee When
considering the 2D cluster state model, it is indeed clear
that it is not one cluster state which forms a universal
resource, but rather the family of all 2D cluster states;
this is most evident in step (ii) in section
\ref{sect_uni_oneway}, where the size of the cluster state
depends on which output state is to be prepared, or which quantum algorithm is to be executed.

Before stating the definition of  universal resource, we
will need the following notation. Let $|\psi\rangle$ be a
multi-qubit state defined on a set of qubits $\{1, \dots,
N\}$, and let $|\phi\rangle$ be an $n$-qubit state, where
$n\leq N$.  We will use the expression $|\psi\rangle
\geq_{\mbox{\tiny LOCC}} |\phi\rangle$ to denote that there
exists a subset of qubits $A\subseteq\{1, \dots, N\}$,
where $|A|=n$, such that the transformation \be
|\psi\rangle \rightarrow |\phi\rangle^A|0\rangle^{\bar
A}\ee (where $\bar A:= \{1, \dots, N\}\setminus A$) is
possible by means of LOCC with unit probability.

We can now state the following definition  of universal
resource, first put forward in  Ref. \cite{Va06}.

\begin{defn}\label{defEDU}{\em \bf (Universal resource):}
A family of states $\Psi =\{|\psi_1\rangle,
|\psi_2\rangle, \dots\}$ is called a universal resource for MQC if, for every $n$, and for every $n$-qubit quantum state $|\phi_{\mbox{\scriptsize
out}}\rangle$, there exists a resource state $|\psi_i\rangle \in \Psi$ such that $|\psi_i\rangle \geq_{\mbox{\tiny LOCC}} |\phi_{\mbox{\scriptsize
out}}\rangle$.
\end{defn}
Thus, we will e.g. say that the family of 2D cluster states \be \Psi_{\mbox{\scriptsize{2D}}} := \{|C_{1\times
1}\rangle, |C_{2\times 2}\rangle, |C_{3\times 3}\rangle, \dots\},
\ee is a universal resource for MQC, as any multi-qubit state can be prepared
(deterministically and exactly) by performing LOCC on 2D cluster state of appropriate size.

The above definition essentially expresses  that a
universal resource needs to be capable of preparing {\em
any} quantum state. Note that we can associate the output
quantum state $|\phi_{\mbox{\scriptsize out}}\rangle$ with
a unitary operation $U$ via \be |\phi_{\mbox{\scriptsize
out}}\rangle:=U|0\rangle^{\otimes n}. \ee This association
makes evident that above definition is concerned with
(CQ-)universal quantum computation, in the sense that a
universal resource can simulate any unitary operation $U$
acting on a fixed input state $|0\rangle^{\otimes n}$.

We now make several remarks regarding the above definition.

\begin{rem} {\it Ignoring efficiency (see also Remark \ref{remark_eff}).---} In
definition \ref{defEDU} the issue of efficiency is
deliberately omitted; for  example, in this definition we
do not put any restrictions on how large a resource state
is allowed to be in order to prepare a desired output
state. The reason for this is, as we will show in section
\ref{sect_ED_criteria}, that several interesting aspects of
universality can already be understood without taking the
additional difficulty of efficiency into account. In
particular, this approach will show its merit when
considering necessary conditions for universality. In
section \ref{sect_ED_criteria} we will present a systematic
approach to obtaining many examples of non-universal
resources---which a fortiori cannot be efficient universal
resources either---without having to consider efficiency
issues. Needless to say that efficiency is of course an
important aspect of the investigation; we will treat this
issue---separately---in the next section, where we provide
a definition for an efficient universal resource. \hfill
$\diamond$ \end{rem}

\begin{rem}\label{rem_LOCC} {\it LOCC versus measurements only.---} In definition \ref{defEDU}  we
slightly extend the framework of one-way quantum
computation and allow for  sequences of {\em arbitrary}
local operations and classical communication, rather than
only local measurements and local Pauli operations. In this
way, the main feature of the one-way model is maintained,
namely that  resource states are processed by local
operations. The resource character of entanglement is
particularly highlighted, as LOCC operations are the most
general operations which cannot increase entanglement.
Hence the resource state includes all the necessary
entanglement which is used up (in part) during the
computation. Of course, it is clear that in realistic
situations one would favor resources which are universal by
local measurements (and some additional local corrections)
only. Note that the examples of universal resources we will
present in section \ref{go}, are all of this form. \hfill
$\diamond$ \end{rem}

\begin{rem} {\it Exact deterministic universality.---} Definition \ref{defEDU}
represents an ideal situation, where we demand that states can be
prepared exactly and with unit probability, i.e., it is a definition
of  exact, deterministic universality. In realistic conditions, it often
suffices that output states can be prepared with an arbitrary high accuracy,
and with a probability which is arbitrarily close to one. This realistic scenario
(called {\em approximate, quasi-deterministic universality}) will not be treated
in the present paper, where we stay with the ideal case in order to outline the
main methods of this study.  The reader is referred to a subsequent article, Ref. \cite{Mo07},  for a treatment of approximate, quasi-deterministic universality. \hfill $\diamond$ \end{rem}

\begin{rem} {\it Fixed vs. random output particles.---} In the above definition for universality, it is assumed (as implicitly also done in Ref. \cite{Va06}) that in all possible branches of the LOCC protocol, the output  state $|\phi_{\mbox{\scriptsize
out}}\rangle$ is prepared on the {\em same} output
particles---see  the discussion preceding definition
\ref{defEDU}. These output systems may in principle be
unknown at the beginning of the LOCC protocol, but need to
be the same for all branches of the protocol. Such a
requirement is natural in the context of CQ-universality,
where a quantum output can be further processed and used
as a resource for some tasks that are specified at a later
stage. One would like to know (favorably already in
advance) on which particles the desired state is going to
be prepared. A situation where this is necessary is e.g.
given by the on-demand preparation of a certain resource
state for distributed security applications that can be
specified by the involved parties at some stage. In this
case, every party holds one of the output particles, and
the preparation of the desired state takes place by LOCC at
some point.

The requirement of fixed output systems can in principle
also be dropped. In this case, one demands instead that the
output state can be generated between some, not previously
specified, subset of particles. The subset of particles may
even be random, depending e.g. on outcomes of measurements
in the LOCC protocol. Note that in the context of
CC-universality it is in fact natural that one does allow
for random output systems (in fact, the
classical output of a ``CC'' cluster state quantum computation is derived by classical
post-processing on the outcomes of measurements on all measured particles, so
it is not clear that this output is localized on the qubits in a strict
sense). In the following, we will {\em
not} consider the possibility of random output systems. We remark, however, that
some of the results we obtain in the next sections
crucially depend on the assumption of a fixed output
systems. We will again comment in remark
\ref{rem_fixed_random2} on the possibility of random output
systems. \hfill $\diamond$ \end{rem}

\subsection{Efficient universality}\label{sect_def_eff}

In classical computation and quantum computation,  {\em
efficiency} is a central issue. Although the above definition
of universality does not take
efficiency issues into account, it is nevertheless useful. As we
have argued to some extent in the previous section, and as
will become more evident in the next one, it is not
necessarily required to consider efficiency. Even when
allowing for (exponential) overheads in temporal and
spatial resources, and for unlimited classical
computational power, one can still rule out large families
of states to be not  universal. Clearly,
if one also considers efficiency, one obtains stronger
criteria for non--universality, and from a practical
perspective only resources which are (in the sense
specified below) efficiently universal are useful.

\subsubsection{Various efficiency issues}

In MQC, the generation of a $n$-qubit state $|\phi_{\rm
out}\rangle$   takes place by performing sequences of
single-qubit measurements, or, more generally, LOCC, on a
resource state of $N\geq n$ qubits. In such a process,
there are  several efficiency issues that one needs to
consider:
\begin{itemize}
\item[(i)] {\it Spatial overhead.---} This refers to  the
required size $N$ of the resource state, since at most
$N-n$ qubits need to be measured in order to generate
$|\phi_{\rm out}\rangle$. We will say that the preparation
of $|\phi_{\rm out}\rangle$ by performing LOCC  on a
resource state is efficient with respect to spatial
overhead if $N={\rm poly}(n)$, i.e., only polynomially many
resource qubits are required. \item[(ii)] {\it Temporal
overhead.---} This refers to the required time steps to
implement the sequence of LOCC. If one is restricted to
projective measurements, then the number of time steps is
at most $N-n$. As shown in Ref. \cite{Ra01, Ra03},  many
measurements can be performed simultaneously; e.g., all
measurements devoted to implement Clifford operations in
the corresponding quantum circuit can be done in a single
time step. For general LOCC, the situation is different.
While it might still be possible to operate on several
qubits simultaneously, sequences of adaptive local
operations of arbitrary length are conceivable. We will say
that the preparation of $|\phi_{\rm out}\rangle$ by
performing LOCC on a resource state is efficient with
respect to temporal overhead, if the number of steps in the
LOCC protocol that cannot be performed in parallel is ${\rm
poly}(n)$, i.e., only polynomially many time steps are
needed. \item[(iii)] {\it Classical side-processing.---}
This is an essential part of the one-way model, where one
needs to keep track of basis changes, and where one has to
decide how measurements (or, more generally, local
operations) are adapted depending on outcomes of previous
measurements. Classical processing may also be required for
other purposes (e.g., when considering error correction).
We will say that the preparation of $|\phi_{\rm
out}\rangle$ by performing LOCC on a resource state is
efficient with respect to classical side-processing if the
overhead for classical computation is polynomially bounded
in space and time. Notice that in the 2D cluster state
model, the time complexity of classical side-processing
only scales as $\log(n)$. \item[(iv)] {\it Description- and
preparation complexity of resource states.---} Throughout
this paper, we will only be interested in whether (families
of) resource states are universal for MQC or not, and we
will {\em not} be concerned with potential difficulties to
{\em describe} the resource states efficiently, or whether
they can be {\em prepared} efficiently \cite{Mo06}. We will
rather assume that the resource states are provided.
\end{itemize}

\begin{rem} {\it Description and preparation complexity.---}
For practical purposes, it may, however, be important to
take the issue (iv) into account. In particular, the possible
efficient preparation of resource states might be crucial
to determine whether MQC could be realized in practice with
a given (family of) resource state(s). By efficient
preparation is meant that a resource state of $N$ qubits
can be prepared with help of a quantum circuit with
poly($N$) elementary gates. Notice, however, that other
ways of preparing resource states are conceivable. For
instance, states that occur naturally as ground states of
certain physical systems may be universal resource states.
If one could find such universal resource states, then the
problem of efficient generation by a quantum
circuit becomes irrelevant. $\quad$ \hfill $\diamond$ \end{rem}

\begin{rem} We (implicitly) assume that the
label $k$ in a family of resource states $\Psi=\{|\psi_k\rangle\}_k$ only
serves as some index to a set of quantum states (most of the times we will have
some regular structure in mind and $k$ is simply related to the size of the structure), and is not used to provide additional computational power. \hfill $\diamond$ \end{rem}

\subsubsection{Definition of efficient
universality}\label{subsect_def_eff}

Having the above efficiency  issues in mind, the following
question arises: {\em which states should be efficiently preparable from an efficient universal
resource?}
Evidently, there are many quantum states $|\phi_{\rm
out}\rangle$ that cannot be generated efficiently even in
the circuit model. As already mentioned in section
\ref{sect_types_uni}, in the definition of efficient
universality for MQC, we will refer to the circuit model
and demand that all states that can be {\em efficiently
prepared in the circuit model should also be preparable
efficiently in an MQC model}. Efficient generation of an
$n$-qubit state $|\phi_{\rm out}\rangle$ with $n$ arbitrary  
in the the circuit model means that there exists a polynomial-size quantum
circuit, i.e., consisting of poly($n$) elementary gates,
which generates the family of these states with $n$ increasing 
from a product state. 
Efficiency in the MQC model refers to efficiency with respect to
spatial and temporal overhead and classical
side-processing, all of which need to be poly($n$). This is
made precise in the following definition.

\begin{defn}\label{defEEDU}{\em \bf (Efficient universal resource)}
A family of states $\Psi$ is called  an efficient universal
resource for MQC if the following is true: for every family
of states $\{|\phi_{\rm out}^{(n)}\rangle\}_{n=1}^{\infty}$
which can be obtained by a poly($n$)-sized family of quantum
circuits, and where $|\phi_{\rm out}^{(n)}\rangle$ is a
state on $n$ qubits, there exists a subfamily \be
\{|\psi_{i_n}\rangle\}_{n=1}^{\infty} \subseteq \Psi,\ee
where $|\psi_{i_n}\rangle$ is a state on at most $N=$
poly($n$) qubits, such that the transformation
\be|\psi_{i_n}\rangle \rightarrow |\phi_{\rm
out}^{(n)}\rangle|0\rangle^{\otimes (N-n)}\ee  is possible
by means of LOCC in poly($n$) time, using classical side
processing that is polynomially bounded in space and time.
\end{defn}
Thus, as already stated in section \ref{sect_uni_oneway},
the set $\Psi_{\rm 2D}$ of $d\times d$ cluster states is an
efficient universal resource. Indeed, as pointed out
earlier, every state which can be prepared by a poly-sized
network, can also be prepared efficiently in the one-way
model [to be precise, one can only meaningfully say that a
\emph{family} of states is generated by a poly-sized
network, rather than a single state. Henceforth, whenever
we refer to ``a state'' which can be prepared by a
poly-sized circuit'', we will mean ``family of states''].

\begin{rem} Equivalently to the above definition, one may define that an efficient universal resource is capable of efficiently preparing any state which can be efficiently prepared in the one-way model.
In this way, one would obtain a definition of efficient
universal resources for MQC which is stated entirely in
terms of measurement-based schemes. \hfill $\diamond$
\end{rem}

\begin{rem} Note that the 2D cluster states can be
described and, more importantly, prepared efficiently (as
is the case for all graph states) \cite{He06, Mh04}.
In particular, the preparation of an $N$-qubit 2D cluster
state requires only $O(N)$ (more precisely: $2N$) phase
gates, most of which can be performed in parallel. The 2D
cluster state can hence be prepared in linear time in the
circuit model. \hfill $\diamond$ \end{rem}

\section{Observations}\label{sect_obs}

In this section, we will present a number of
straightforward observations which will turn out to be
crucial to establish both necessary and sufficient criteria
for (efficient) universality.

\subsection{Universality and the 2D cluster states}
We start with a key observation regarding universal resources which has already been presented in Ref. \cite{Va06}.

\begin{obs}\label{obs1}
{\it A set of states $\Psi$ is a universal resource if and
only if  all 2D cluster states $|C_{d\times d}\rangle$ (for
all $d$) can be prepared (deterministically and exactly)
from the set $\Psi$ by LOCC.}
\end{obs}
Necessity of the condition follows  from the fact that a
universal resource should be capable of  preparing {\em
any} quantum state, in particular an arbitrary 2D cluster
state. Sufficiency follows from the fact that once a 2D
cluster state of arbitrary size can be created, one can use
the one-way model for quantum computation to generate {\em
any} quantum state by LOCC \cite{Ra01, Ra03}.

Note that a direct consequence of this observation is that
every CQ-universal resource can immediately be made
QQ-universal by reducing it to the 2D cluster state model.
This can be seen as follows. Due to the fact that from any
universal resource a 2D cluster state can be generated, a
possible way to perform an arbitrary (CQ) quantum
computation (i.e., to generate an $n$-qubit state
$|\phi_{\rm out}\rangle = U |0\rangle$) is to first
generate a sufficiently large 2D cluster state, and then
continue with the {\em same} LOCC protocol as in the
one-way model.  In particular, this allows one to use the
same methods and techniques as in the one-way model for MQC
to process classical inputs. What is more, in this way any
CQ-universal resource $\Psi$ can immediately be made
QQ-universal by supplying $\Psi$ with the same input
coupler (i.e., an additional resource in the form of a
simple round of $n$ Bell measurements or controlled phase
gates) as in the one-way model.

\begin{obs}\label{obs1'}
{\it Every CQ-universal resource can be made QQ-universal by allowing quantum inputs to be coupled in to the resource by the same input coupler as in the one-way model, provided e.g. by $n$ Bell measurements or phase gates.}
\end{obs}

This observation illustrates in particular that there is no
severe restriction in the fact that MQC schemes (within the
framework ``resource state plus LOCC'') can only be
CQ-universal, since only a small step is required to go
from CQ-universality to the most general notion of
QQ-universality.

\subsection{Efficient universality and the 2D cluster
states}\label{subsect_obs2}

An observation for efficient universality is the following.

\begin{obs}\label{obs2}
{\it A set of states $\Psi$ is an efficient universal resource if and only if all 2D cluster states $|C_{d\times d}\rangle$ (for all $d$) can be efficiently prepared from the set $\Psi$ by LOCC, i.e., with polynomial spatial and temporal overhead, and with polynomial classical side-processing.}
\end{obs}
Necessity of the condition follows from the fact that an efficient universal resource  must  be capable of preparing the 2D cluster states efficiently, since these states can be prepared efficiently in the circuit model. Sufficiency follows from the fact that, once a 2D cluster state of arbitrary size can be created efficiently, one can use the one-way model to efficiently generate {\em any} quantum state that can be efficiently generated in the circuit model (as discussed in the previous section).

Again, this observation leads to one possible way of
performing the computation by generating first a 2D cluster
state and then using the corresponding LOCC protocol of the
one-way model. Similar as in the case where efficiency was
not considered, it follows that any efficient CQ-universal
resource is also efficiently QQ-universal when allowing for same
supplementary resource (``input coupler'') as in the one-way
model for MQC.

\subsection{Equivalence of universal graph states}

The next observation is concerned with a particular family
of potential resource states, the graph states \cite{He06}.
Consider a family $\Psi$ where all members are graph
states, and consider the case where we are interested only
in preparing a specific quantum state. In  principle it is
conceivable that such a specific task can be performed more
efficiently for certain graph resource states than for the
2D cluster state. The following observation, however, shows
that this is not the case.

\begin{obs}\label{obs3}
{\it Any quantum state that can be prepared from a graph state resource with polynomial temporal and spatial overhead by means of LOCC, can also be prepared from a 2D cluster state with polynomial temporal and spatial overhead.}
\end{obs}
This observation implies that one cannot hope for an (exponential) speedup, even to prepare certain specific states $|\phi_{\rm out}\rangle$ (or, equivalently, to perform some specific  unitary operation, i.e., algorithm) within the measurement-based model if the resource states are general graph states. The one-way model based on the 2D cluster state is already optimal in this respect. The observation immediately follows from the fact that {\em any} graph state of $n$ qubits can be prepared efficiently in the circuit model with at most $n(n-1)/2$ two-qubit phase gates \cite{He06}.  This implies that any graph state can also be prepared efficiently, i.e. with polynomial overhead in spatial and temporal resources, and in classical side--processing, from a 2D cluster state.

Note that the reverse of this observation is not
necessarily true. One can imagine families of graph states
that are designed in such a way that they include a useful
part (e.g., some form of 2D cluster state of restricted
size), but in addition there are a large number of useless
``dummy''-particles. If for a fixed number of qubits $N$ the
number of dummy-particles is exponentially larger than the
useful part of size $n$, then one will always have an
exponential overhead in spatial resources, e.g., to prepare
a $n$-qubit cluster state.

Nevertheless, if we restrict ourselves to efficient
universal graph state resources, one  finds that all such
resources are equivalent, as they can be obtained from each
other with polynomial spatial and temporal overhead.
\begin{obs}\label{obs4}
{\it All efficient universal graph state resources can
efficiently be obtained from each other by means of LOCC.}
\end{obs}

This property follows from the fact that all graph  states
can be prepared efficiently in the circuit model.

\section{Criteria for universality and no-go results}\label{sect_ED_criteria}

In this section, we  formulate several requirements which
every universal resource must meet. These necessary
conditions are all formulated in terms of entanglement---as
measured by certain appropriate entanglement
measures---which needs to be present in every universal
resource, hence emphasizing the role of entanglement as a
resource for MQC.

In section \ref{sect_I_strategy}, we give an outline of the
general strategy which will be used to formulate criteria
for universality. After this, in sections \ref{sect_I_ewd}
through \ref{sect_I_sm}, we apply this approach. We
formulate several necessary conditions for  universality,
and give numerous examples of resources which do not comply
with these criteria, and which hence cannot be universal.

\subsection{Type II entanglement monotones}\label{sect_I_strategy}

In this section, the general strategy to obtain necessary
conditions for
  universality is outlined. We emphasize that, at this point, we are only
concerned with universality, and not with efficient
universality. The additional aspect of efficiency will be
treated in section \ref{sect_I_eff}.

First we need some definitions and notation. Let
$E(|\psi\rangle)$ be a functional defined for all $n$-qubit
states $|\psi\rangle$, for all natural numbers $n$. We
denote by $E^*$ the supremal value of $E$, when the
supremum is taken over all possible $n$-qubit states, for
all $n$ (the case $E^* = \infty$ is allowed): \be E^*:=
\sup_{\mbox{\scriptsize{all }}|\psi\rangle}
E(|\psi\rangle). \ee Furthermore, letting
$\Psi=\{|\psi_1\rangle, |\psi_2\rangle, \dots\}$ be an
arbitrary resource, we define $E(\Psi)$ to be the supremal
value of $E$, when the supremum is taken over all states in
the resource $\Psi$, i.e., \be E(\Psi):=
\sup_{|\psi_i\rangle\in\Psi} E(|\psi_i\rangle)\ee We are
now interested in functionals $E$ satisfying the following
property:

\begin{itemize}
\item[(P1)] $E(|\psi\rangle)\geq E(|\psi'\rangle)$ whenever
$|\psi\rangle\geq_{\mbox{\tiny LOCC}}|\psi'\rangle$, for
every $N$-qubit state $|\psi\rangle$ and $n$-qubit state
$|\psi'\rangle$, and for every $N, n$ with $N\geq n$.
\end{itemize}
Property (P1) states that $E$ is a measure which is
non-increasing under \emph{deterministic} LOCC interconversion between (pure)
quantum states. Measures satisfying this property are similar to entanglement monotones \cite{Vi00, Pl07}, which are defined to be non-increasing \emph{on average} under LOCC. A functional satisfying property (P1) will henceforth be called a \emph{type II entanglement monotone} in order to distinguish it from the standard notion of an entanglement monotone; the latter will be called a type I monotone. We refer the reader to the end of this section for a more thorough discussion concerning type I and type II monotones.

We can now formulate the following simple
result, which will be central to our analysis.
\begin{thm}\label{thm_P1}
Let $\Psi$ be universal resource   and let $E$ be a
type II entanglement monotone. Then $E(\Psi) = E^*.$
\end{thm}
{\it Proof:} As $\Psi=\{|\psi_1\rangle, |\psi_2\rangle,
\dots\}$ is universal resource,  there exists, for every
multi-qubit state $|\phi\rangle$, a state
$|\psi_{i}\rangle\in \Psi$ (dependent on $|\phi\rangle$)
such that $|\psi_{i}\rangle \geq_{\mbox{\tiny
LOCC}}|\phi\rangle$. This implies that $E(|\psi_{i}\rangle)
\geq E(|\phi\rangle)$, since $E$ is a type  II monotone.
This immediately implies that \be E(\Psi)\geq
\sup_{|\phi\rangle} E(|\phi\rangle) = E^*.\ee Using the
definition of $E^*$ as a supremum then implies that
$E(\Psi)=E^*$, yielding the desired result. \finpr

As an  immediate corollary to theorem \ref{thm_P1} and the
 universality of the 2D cluster states,
we find the following:
\begin{cor}
Let $E$ be a type II monotone and let $\Psi_{2D}$ be the
family of 2D cluster states. Then $E(\Psi_{2D}) = E^*.$
\end{cor}
Theorem \ref{thm_P1} states that all universal resources---as  e.g. the family of 2D cluster states---are
\emph{maximally entangled}, in the sense that every type II monotone
$E$  has to reach its supremum on every
universal resource.  Note that this implies that all universal resources  have many entanglement features in common,
namely all features quantified by type II monotones.
In order to investigate the entanglement present in general
universal resources,  it therefore suffices to single
out one such resource---say, the family of 2D cluster
states---and study its entanglement features as measured by
the functionals $E$. In this sense, the entanglement
present in the 2D cluster states is representative for the
entanglement present in all universal resources.  We
emphasize that the 2D cluster states a priori do \emph{not}
play a distinguished role in this context, as they can be
replaced by any other universal resource.  However, they
usually form an appropriate test-bed, as many of their
properties have already been established (see e.g. Ref.
\cite{He06}), and since the fact that these states are graph
states often allows one to use powerful tools from the
stabilizer formalism \cite{Go97} to investigate their properties.

Evidently, in order to show that a given family of states is
\emph{not} universal,  it suffices to
find a suitable measure $E$ such that theorem \ref{thm_P1}
is violated:
\begin{cor}\label{cor_criterion}
Let $\Psi$ be a resource, and suppose that there exists a type II monotone
$E$, such that $E(\Psi)< E^*.$
Then $\Psi$ cannot be a universal resource.
\end{cor}
This corollary captures the  main strategy which will be
adopted in the following in order to obtain necessary
conditions for  universality. To arrive
at a criterion, one simply has to specify a functional $E$,
prove that it is a type II monotone, and finally compute its
supremum $E^*$.  Note that, in order to compute $E^*$, it
suffices to compute the supremal value of $E$ on the 2D
cluster states, or any other universal resource. This is
sometimes (but not always) more convenient than considering
the definition of $E^*$ as a supremum over all states, as
we will see below.

Note that the definition of a type II monotone is related
to, but slightly different from, the definition of a (type
I) entanglement monotone as introduced in Ref. \cite{Vi00}. A (type
I) entanglement monotone $M$ is a functional  defined on
the set of $n$-qubit states such that $M$ decreases
\emph{on average} under LOCC (where usually LOCC between
states of the same system size are considered). That is, if
an LOCC protocol is executed on an $n$-qubit input state
$|\psi\rangle$, leading to $n$-qubit output states
$\{|\psi_i\rangle\}$ in the different branches of the
protocol, with probabilities $\{p_i\}$, then a monotone $M$
by definition satisfies \be\label{monotone} M(|\psi\rangle)
\geq \sum_{i} p_i M(|\psi_i\rangle).\ee This is slightly
different from the requirement (P1) for two reasons. First,
in (P1) deterministic LOCC conversions between two states
$|\psi\rangle$ and $|\psi'\rangle$ are considered, rather
than protocols with different output states occurring with
certain probabilities---hence, in (P1) there is no
averaging as in (\ref{monotone}). Second, in (P1) states of
possibly different system sizes ($N$ and $n$) are compared.

Despite of these  differences, it is clear that there
exists a large overlap between type I and type II
monotones. In particular, note that, for every $n$-qubit
type I monotone $M$, one has $M(|\psi\rangle)\geq
M(|\psi'\rangle)$ for any pair of $n$-qubit states such
that $|\psi\rangle$ can deterministically be transformed
into $|\psi'\rangle$ by means of LOCC (note again that here
states of the \emph{same} system size are considered).
Therefore, it comes a no surprise that several type I
monotones are also type II monotones. The following result
is easily verified.

\begin{thm}\label{thm_monotone}
Let $E(|\psi\rangle)$ be a functional defined for all
$n$-qubit states $|\psi\rangle$, for all natural numbers
$n$. Suppose that the following statements hold:
\begin{itemize}
\item[(i)] $E$ is a type I entanglement monotone---i.e.,
$E$ is non-increasing on average under LOCC, when states of
the same system size are considered; \item[(ii)]
$E(|\psi\rangle|0\rangle) = E(|\psi\rangle)$ for every
multi-qubit state $|\psi\rangle$---i.e., $E$ is invariant
under the addition of an uncorrelated one-qubit state
(constituting an extra party).
\end{itemize}
Then $E$ is a type II monotone.
\end{thm}
Thus, all type I entanglement monotones  which are
invariant under the adding of uncorrelated single-qubit
states, are also type II monotones, and thus are suitable
measures to obtain necessary conditions for  universality.
We will encounter examples of this below.

On the other hand, we emphasize that several type I entanglement
monotones do not meet requirement (ii) in theorem
\ref{thm_monotone}, and are therefore not useful to obtain
necessary conditions for universality.
Examples of such measures can straightforwardly be obtained
as follows. Consider any type I monotone $\bar E$ which is defined by \emph{averaging} a bipartite monotone
$E$ over all bipartitions of the system. It
is clear that $\bar E$ then does not satisfy (ii), as appending
one-qubit parties which are disentangled from the rest of
the system will decrease the average entanglement.
Therefore, the class of such 'averaging' type I monotones $\bar E$
do not give rise to suitable necessary conditions for exact
deterministic universality. Other examples can easily be
given.

Finally, we note that there exist type II entanglement monotones which are not type I monotones. We refer to e.g. the next section, where it is shown that the entropic entanglement width is a type II but not a type I monotone.

In the following sections (i.e., sections \ref{sect_I_ewd}
to \ref{sect_I_sm}), we  will follow the approach
formulated after corollary \ref{cor_criterion} by applying
it to several measures $E$. We will in particular consider
the \emph{entanglement width} measures, several measures
related to \emph{localizable entanglement}, the
\emph{geometric measure} of entanglement and the
\emph{Schmidt measure}. Necessary conditions for
universality will be obtained by showing that these
measures are all type II monotones. Moreover, we will
encounter examples of resources $\Psi$ which violate at
least one of the resulting criteria, such that these
resources cannot be universal.

\subsection{Entanglement width}\label{sect_I_ewd}
In this section, we discuss two measures,  introduced and
investigated in Refs. \cite{Va06, Va06a}, namely the
\emph{entropic entanglement width} and the
\emph{Schmidt-rank width}. We will show that these measures
are type II monotones, and hence give rise to criteria for
universality. We will subsequently use these criteria to
obtain several examples of non-universal resources.

\subsubsection{Definitions}

The entropic entanglement width $E_{\mbox{\scriptsize
wd}}(|\psi\rangle)$ of an multi--party state $|\psi\rangle$
is an entanglement measure introduced in Ref. \cite{Va06}.
This measure computes the minimal bipartite entanglement
entropy in the state $|\psi\rangle$, where the minimum is
taken over specific classes of bipartitions of the system.
The precise definition is the following.

\begin{figure}
\hspace{4.5cm}\includegraphics[width=0.50\textwidth]{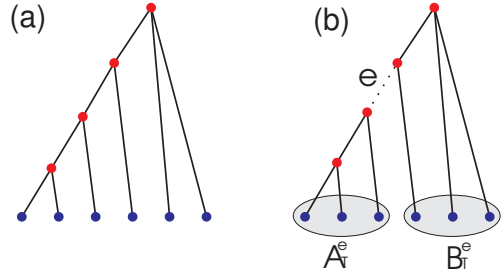}
\caption[]{\label{subcubic} (a) Example of a subcubic tree
$T$ with six leaves (indicated in blue). (b) Tree
$T\setminus e$ obtained by removing edge $e$ and induced
bipartition $(A_{T}^e, B_{T}^e)$.}
\end{figure}

Let $|\psi\rangle$ be an $N$-qubit pure state. A
\emph{tree} is a graph with no cycles. Let $T$ be a
\emph{subcubic} tree, which is a tree such that every
vertex has exactly 1 or 3 incident edges. The vertices
which are incident with exactly one edge are called the
\emph{leaves} of the tree. We consider trees $T$ with
exactly $N$ leaves $V:=\{1, \dots, N\}$, which are
identified with the $N$ qubits of the system. Letting
$e=\{i, j\}$ be an arbitrary edge of $T$, we denote by
$T\setminus e$ the graph obtained by deleting the edge $e$
from $T$. The graph $T\setminus e$ then consists of exactly
two connected components (see Fig. \ref{subcubic}), which
naturally induce a bipartition $(A_{T}^e, B_{T}^e)$ of the
set of qubits $V$. We denote the bipartite entanglement
entropy of $|\psi\rangle$ with respect to the bipartition
$(A_{T}^e, B_{T}^e)$ by $E_{A_{T}^e,
B_{T}^e}(|\psi\rangle)$, where $E_{A,B}(|\psi\rangle)=
-\tr(\rho_A \log_2 \rho_A)$ with
$\rho_A=\tr_B(|\psi\rangle\langle\psi|)$. The
\emph{entropic entanglement width} of the state
$|\psi\rangle$ is now defined by \be E_{\mbox{\scriptsize
wd}}(|\psi\rangle):= \min_T\ \max_{e\in T}\ E_{A_{T}^e,
B_{T}^e}(|\psi\rangle), \ee where the minimization is taken
over all subcubic trees $T$ with $N$ leaves, which are
identified with the $N$ parties in the system. Thus, for a
given tree $T$ we consider the maximum, over all edges in
$T$, of the quantity $\ E_{A_{T}^e,
B_{T}^e}(|\psi\rangle)$; then the minimum, over all
subcubic trees $T$,  of such maxima is computed.

Similarly, one can use the Schmidt rank, i.e., the number
of non--zero Schmidt coefficients, instead of the bipartite
entropy of entanglement as a basic measure. One then obtains
the \emph{Schmidt--rank width}, or \emph{$\chi$--width},
denoted by $\chi_{\mbox{\scriptsize{wd}}}(|\psi\rangle)$
\cite{Va06a}. The precise definition is the following.
Let $\chi_{A^e_T, B^e_T}(|\psi\rangle)$ denote the
number of non-zero Schmidt coefficients of $|\psi\rangle$
with respect to a bipartition $(A_{T}^e, B_{T}^e)$ of $V$
as defined above, i.e. $\chi_{A,B}(|\psi\rangle) = {\rm rank}(\rho_A)$.
The $\chi$--width of the state
$|\psi\rangle$ is defined by \be \label{chiwidth}
\chi_{\mbox{\scriptsize{wd}}} (|\psi\rangle):=
\min_{T}\max_{e\in T} \log_2 \chi_{A^e_T,
B^e_T}(|\psi\rangle). \ee

Note that, since the inequality \be\log_2 \chi_{A,
B}(|\psi\rangle)\geq E_{A, B}(|\psi\rangle)\ee holds for
any bipartition $(A, B)$ of the system and for any state
$|\psi\rangle$, we have \be\label{ineq_width}
\chi_{\mbox{\scriptsize{wd}}}(|\psi\rangle)  \geq
E_{\mbox{\scriptsize wd}}(|\psi\rangle).\ee The
entanglement width measures are extensively studied in
Refs. \cite{Va06, Va06a}, to which we refer the interested
reader for more details. In particular, it was shown that
the Schmidt-rank width can be given a natural
interpretation, as this measure quantifies the optimal
description of a state in terms of a tree tensor network.

\subsubsection{Formulation of the universality criterion}\label{sect_criterion_width}

Next we prove that $E_{\mbox{\scriptsize wd}}$ and
$\chi_{\mbox{\scriptsize wd}}$ are type II monotones. First
we focus on the entropic entanglement width.
\begin{thm}\label{thm_P1_width}
The entropic entanglement width $E_{\mbox{\scriptsize
wd}}(|\psi\rangle)$ is a type II monotone.
\end{thm}
{\it Proof:} Let $|\phi\rangle$ and $|\phi'\rangle$ be two
$n$-qubit states, such that $|\phi\rangle$ is
(deterministically) convertible by LOCC into
$|\phi'\rangle$. Let $T_0$ be a subcubic tree such that \be
\max_{e\in T_0}\ E_{A_{T_0}^e, B_{T_0}^e}(|\phi\rangle) =
E_{\mbox{\scriptsize{wd}}}(|\phi\rangle).\ee Moreover, let
$e_0$ be an edge of $T_0$ such that \be E_{A_{T_0}^{e_0},
B_{T_0}^{e_0}}(|\phi'\rangle) = \max_{e\in T_0}\
E_{A_{T_0}^e, B_{T_0}^e}(|\phi'\rangle).\ee We then have:
\be E_{\mbox{\scriptsize{wd}}}(|\phi\rangle)&=&\max_{e\in
T_0}\ E_{A_{T_0}^e, B_{T_0}^e}(|\phi\rangle)\nonumber\\
&\geq& E_{A_{T_0}^{e_0}, B_{T_0}^{e_0}}(|\phi\rangle) \\
&\geq& E_{A_{T_0}^{e_0}, B_{T_0}^{e_0}}(|\phi'\rangle)\\
&=&\max_{e\in T_0}\ E_{A_{T_0}^e,
B_{T_0}^e}(|\phi'\rangle)\\ &\geq&
E_{\mbox{\scriptsize{wd}}}(|\phi'\rangle).\ee Further,
suppose that $|\psi\rangle$ and $|\psi'\rangle$ be
$N$-qubit and $n$-qubit states, respectively, such that
$|\psi\rangle \geq_{\mbox{\tiny LOCC}}|\psi'\rangle$. Then
the above implies that  \be
E_{\mbox{\scriptsize{wd}}}(|\psi\rangle) \geq
E_{\mbox{\scriptsize{wd}}}(|\psi'\rangle
|0\rangle^{N-n}).\ee  The result is obtained if
$E_{\mbox{\scriptsize{wd}}}(|\psi\rangle |0\rangle) =
E_{\mbox{\scriptsize{wd}}}(|\psi\rangle)$ for all states
$|\psi\rangle$. This is essentially implied by the
following property. Suppose that $|\psi\rangle$ is an
$N$-qubit state defined on a set of qubits $V:=\{1, \dots,
N\}$, and suppose that $|\psi\rangle|0\rangle$ is defined
on the set $V':= \{1, \dots, N, N+1\}= V \cup \{N+1\}$.
Letting $(A, B)$ be an arbitrary bipartition of $V$ and
writing $A':= A\cup \{N+1\}$, we have $E_{A,
B}(|\psi\rangle) = E_{A', B}(|\psi\rangle|0\rangle).$ This
property can straightforwardly be used to show that the
entropic entanglement width is the same for the states
$|\psi\rangle$ and $|\psi\rangle|0\rangle$. \finpr

Notice, however, that the entropic entanglement  width is
{\em not a type I entanglement monotone}. Despite the fact
that (P1) is satisfied, this measure can increase on
average under LOCC. This is most easily verified by
considering the example of a three-qubit W state
\cite{Du00}, from which maximally entangled pairs shared
between random pairs of parties can be generated with
probability arbitrarily close to one \cite{Fo06}. While the
entropic entanglement width of the W state is smaller than
one, the entanglement width of the output Bell pairs is
equal to one.

An argument similar to the proof of theorem
\ref{thm_P1_width} can be used to prove that the
Schmidt-rank width also satisfies (P1), i.e., this measure is a type II monotone. In fact, as is e.g.
shown in Ref. \cite{Mi03}, the Schmidt rank is non-increasing under
stochastic local operations and classical communication
(SLOCC). It follows that in contrast to the entropic
entanglement width, the Schmidt rank width is a type I
entanglement monotone, and even satisfies a stronger
condition:

\begin{thm}\label{thm_P1_chi} Consider and $N$-qubit state $|\psi\rangle$ and an $n$-qubit state
$|\psi'\rangle$ with $N\geq n$, such that $|\psi\rangle$
can be transformed into $|\psi'\rangle|0\rangle^{N-n}$ by
SLOCC with some non-zero probability. Then
$\chi_{\mbox{\scriptsize{wd}}}(|\psi\rangle)\geq
\chi_{\mbox{\scriptsize{wd}}}(|\psi'\rangle)$. As a
corollary, the Schmidt-rank width is a type I and type II monotone.
\end{thm}
In order to formulate the criteria for universality
associated to the entanglement width measures, we now
compute the supremal values of these measures. We will
prove that both of these suprema are unbounded, i.e.,
$E_{\mbox{\scriptsize wd}}^* = \chi_{\mbox{\scriptsize
wd}}^* = \infty$.
To obtain these results, it will suffice
to show that the entropic entanglement width of the 2D
cluster states is unbounded \footnote{In fact, any randomly chosen state has the property that the entropy of any reduced density operator is almost maximal (see Ref. \cite{Pa93}), and hence also the entropic entanglement width is unbounded.}.
Notice that the infinity of
$E_{\mbox{\scriptsize wd}}$ on the 2D cluster states implies the infinity of $E_{\mbox{\scriptsize
wd}}^*$. Moreover, the latter implies infinity of
$\chi_{\mbox{\scriptsize wd}}^*$ via equation
(\ref{ineq_width}).

In order to prove that 2D cluster states have an unbounded
entropic entanglement width, we need to evaluate this
measure on graph states. Here we use a result established
in Refs. \cite{Va06, Va06a}, stating that the entropic
entanglement width of a graph state is equal to the
\emph{rank width} of the underlying graph. The rank width
is a graph invariant defined in Ref. \cite{Oum}. One then
uses the property that the rank width of the $(2l+1)\times
(2l+1)$ grid graph is lower bounded by $l-1$
\cite{Oum_private}, showing that this measure is indeed
unbounded in the limit of infinitely many qubits. This
shows that $E_{\mbox{\scriptsize wd}}^* =
\chi_{\mbox{\scriptsize wd}}^* = \infty$.

Combining the above argument with theorems \ref{thm_P1},
\ref{thm_P1_width} and \ref{thm_P1_chi} we arrive at the
following criterion for universality.

\begin{thm}\label{thm_ED_ewd}
Let $\Psi$ be a  universal resource. Then
$E_{\mbox{\scriptsize wd}}(\Psi) = \chi_{\mbox{\scriptsize
wd}}(\Psi) = \infty.$ Hence, any resource with a bounded
$E_{\mbox{\scriptsize wd}}$ or $\chi_{\mbox{\scriptsize
wd}}$ cannot be universal.
\end{thm}
Although the definitions of the entanglement width measures
involve highly nontrivial combinatorial optimization
problems, these measures can efficiently be evaluated (or
approximated) on several families of states---and hence the
corresponding criteria for  universality
can be employed. In particular, the entanglement width
criteria prove to be very powerful to investigate the
universality of graph state resources. This is illustrated
in the following example.

{\it Example.---} It was shown in Refs. \cite{Va06, Va06a}
that both the entropic entanglement width and the
Schmidt-rank width of an arbitrary graph state  coincide
with the rank width of the underlying graph. Together with
theorem \ref{thm_ED_ewd}, this implies that every graph
state resource where the rank width of the underlying
graphs is \emph{bounded}, cannot be a universal resource.
This criterion leads to many examples of graph state
resources which are not universal,  as several examples of
families of graphs are known where the rank width measure
is bounded---in fact, efficient algorithms exists to
compute (or approximate) the rank width of any graph
\cite{Oum}. Examples of graphs of bounded rank width,
giving rise to non-universal resources, have been given in
previous work, see Refs. \cite{Va06, Va06a}. These examples
include

\begin{itemize}
\item tree graphs \item cycle graphs,\item  co-graphs,
\item graphs locally equivalent to trees, \item
distance-hereditary graphs, \item graphs of bounded tree
width, \item graphs of bounded clique width, \item ...
\end{itemize}
We refer to the graph theory literature for definitions.
Note that two interesting examples of non-universal
resources are the linear cluster states (which are
instances of tree graphs) and the GHZ states (corresponding to complete graphs), or any family
of $k\times l$ lattices where $k$ is constant (and $l$
grows with the system size). \hfill $\diamond$

A second application of theorem \ref{thm_ED_ewd} is
obtained by considering ground states of strongly
correlated spin systems associated to a one-dimensional
geometry.

{\it Example.---}  Consider ground states of strongly
correlated spin systems with nearest-neighbor interaction
associated to a one-dimensional geometry. When such systems
are in the \emph{non-critical} phase, one typically finds
that the bipartite entropy of entanglement between a
subchain of spins and the rest of the chain, is bounded
from above by a constant. See e.g. Ref.
\cite{Vi03}. For all ground states where
this property holds, one can easily show that the entropic
entanglement width is bounded. To do so, let
$|\psi_N\rangle$ be such a ground state of a system of $N$
particles, and consider the subcubic tree $T$ depicted in
Fig. \ref{subcubic}, where the leaves $1$ to $N$ correspond
to the natural linear ordering of the system. As all
bipartitions $(A_T^e, B_T^e)$  are such that some connected
subchain of particles $\{a, a+1, \dots, a+k\}$ is separated
from the rest of the system, one finds that the quantity
\be\label{upper} \max_{e\in T} E_{A_T^e,
B_T^e}(|\psi\rangle)\ee is bounded from above by a constant
(independent of $N$). As (\ref{upper}) is by definition an
upper bound to the entropic entanglement width, one finds
that this measure is bounded on the resource
$\Psi:=\{|\psi_N\rangle\}_N$. This implies that $\Psi$
cannot be universal.  We have therefore found that any
resource of ground states of (non-critical) 1D
spin systems where the block-wise entanglement entropy is
bounded, cannot be a universal resource.

Below (see section
\ref{sect_I_eff}) we will see that a similar result
typically holds for ground states of \emph{critical} 1D
systems, and we will prove that such ground states cannot
be \emph{efficient} universal resources.  \hfill $\diamond$

\subsection{Localizable entanglement}\label{sect_I_le}

Next we discuss a second class of measures which gives rise
to criteria for universality. These measures will be
centered around the possibility or impossibility of
preparing  Bell states by performing LOCC on resource
states.

Consider the following simple measure
$E_{\mbox{\scriptsize{Bell}}}$: for an arbitrary $N$-qubit
state $|\psi\rangle$, define
$E_{\mbox{\scriptsize{Bell}}}(|\psi\rangle):=1$ if \be
|\psi\rangle \geq_{\mbox{\tiny LOCC}}
\frac{1}{\sqrt{2}}(|00\rangle + |11\rangle),\ee i.e., if it
is possible to deterministically create a Bell pair between
a predefined pair of qubits in the system, and
$E_{\mbox{\scriptsize{Bell}}}(|\psi\rangle):=0$ if this is
not possible. Clearly, one has
$E_{\mbox{\scriptsize{Bell}}}^* = 1$, and
$E_{\mbox{\scriptsize{Bell}}}$ is a type II monotone.
Therefore, every universal resource  $\Psi$ must satisfy
$E_{\mbox{\scriptsize{Bell}}}(\Psi)=1$. This is nothing but
stating that a universal resource---i.e., a  resource
capable of preparing \emph{arbitrary} states---must be able
to prepare Bell states, which is a trivial observation.
Nevertheless, this this simple criterion allows one to
conclude that many resources are not universal. For
instance,  one can consider the following example.

{\it Example.---} The family of W states
$\{|W_N\rangle\}_N$ is not universal, as it has been shown
that Bell state cannot be prepared deterministically by
performing LOCC on W states \cite{Fo06}.  Here we have used
the definition \be |W_N\rangle:=
\frac{1}{\sqrt{N}}\sum_{i=1}^N |e_{N, i}\rangle,\ee where
$|e_{N, i}\rangle$ is defined to be the $N$-qubit
computational basis state with a $|1\rangle$-state on the
$i$th position in the tensor product, and $|0\rangle$
everywhere else (for example $|e_{3,1}\rangle =
|100\rangle$, $|e_{3,2}\rangle = |010\rangle$, ...).\hfill
$\diamond$

Using this idea, a slightly more involved but substantially
more powerful criterion can be obtained as follows. Let
$|\psi\rangle$ be a state on $N$ qubits $V:=\{1, \dots,
N\}$ and let $\alpha$ and $\beta$ be two qubits in the
system. Define \be E^{\alpha,
\beta}_{\mbox{\scriptsize{Bell}}}(|\psi\rangle):=1\ee if it
is possible to deterministically create a Bell pair on the
qubits $\alpha$ and $\beta$ by performing LOCC on
$|\psi\rangle$, and this measure is zero otherwise. Then
define the measure ${\cal
N}_{\mbox{\scriptsize{LE}}}(|\psi\rangle)$ to be the
maximal size $|A|$ of a subset of qubits $A\subseteq V$
such that $E^{\alpha,
\beta}_{\mbox{\scriptsize{Bell}}}(|\psi\rangle)=1 \mbox{
for all } \alpha \mbox{ and } \beta\mbox{ in } A \mbox{
where }\alpha\neq\beta$. Thus, ${\cal
N}_{\mbox{\scriptsize{LE}}}(|\psi\rangle)$ is the largest
size of a subset of qubits, such that a Bell state can be
created deterministically between any two qubits in this
subset, by performing LOCC on $|\psi\rangle$.

It is again clear from the definition that the measure
${\cal N}_{\mbox{\scriptsize{LE}}}$ is a type II monotone.
To compute the supremum ${\cal
N}_{\mbox{\scriptsize{LE}}}^*$, we again consider the 2D
cluster states. It is well known that when a system of
qubits is in a 2D cluster state, then a Bell pair can be
created between any pair of qubits by LOCC \cite{Br01,
He06}. Therefore, ${\cal
N}_{\mbox{\scriptsize{LE}}}(|C_{d\times d}\rangle) = d^2$,
which shows that ${\cal N}_{\mbox{\scriptsize{LE}}}^* =
\infty$. This leads to the following result.

\begin{thm}\label{thm_ED_le}
Let $\Psi$ be a  universal resource.  Then
${\cal N}_{\mbox{\scriptsize{LE}}}(\Psi) = \infty$.
\end{thm}
In order to obtain examples of no-go
results, we again
turn to ground states of strongly correlated spin systems.

{\it Example.---}  In the context of spin systems, the
entanglement measure \emph{localizable entanglement},
introduced in Ref. \cite{Ve_loc04}, is often considered.
This measure, which will be connected to ${\cal
N}_{\mbox{\scriptsize{LE}}}$, is defined as follows. Let
$|\psi\rangle$ be an $N$-qubit state, and let $\alpha$ and
$\beta$ denote two arbitrary qubits in the system. The
localizable entanglement between these qubits, denoted by
$L^{\alpha\beta}(|\psi\rangle)$, is defined to be maximal
entanglement (measured by the concurrence) of the subsystem
of qubits $\{\alpha, \beta\}$, which can be created
\emph{on average} by performing LOCC on the other qubits in
the system. This quantity is an entanglement monotone for
$2 \times 2 \times l$ systems \cite{Go05} and also fulfills
property (P1). Note that $L^{\alpha\beta}$ is maximal
(i.e., equal to one) if and only if a Bell pair can
deterministically be created between the qubits $\alpha$
and $\beta$ by LOCC---in other words,
$L^{\alpha\beta}(|\psi\rangle)=1$ if and only if
$E^{\alpha,
\beta}_{\mbox{\scriptsize{Bell}}}(|\psi\rangle)=1$. In
ground states of strongly correlated spin systems which are
organized according to some geometry (such as a
$d$-dimensional lattice), $L^{\alpha\beta}$ is typically a
decreasing function of the distance between the spins
$\alpha$ and $\beta$ \cite{Ve_loc04, decay_LE}. Note that,
whenever such a decay law of the localizable entanglement
exists, that the measure ${\cal
N}_{\mbox{\scriptsize{LE}}}$ must be bounded. This implies
that, if the ground state of a spin system exhibits a decay
of the localizable entanglement with the distance between
spins, then this resource cannot be universal.\hfill
 $\diamond$

\begin{rem} {\it Extension to approximate quasi-deterministic universality.---}
We note that, contrary to the results obtained for the
entanglement width measures, one cannot extend theorem
\ref{thm_ED_le} to the case of approximate and/or
quasi-deterministic universality (see our upcoming work
\cite{Mo07}).  Hence, it will be restricted to the case of exact
deterministic universality as studied in the present
paper.\hfill $\diamond$
\end{rem}

\begin{rem}\label{rem_fixed_random2} {\it Fixed  versus random output particles.---} Notice that the above arguments are based on the assumption (as already discussed earlier) that the output particles belong to a fixed set $A$ in all branches of the LOCC protocol. If one would allow to generate the desired output state between {\em some}, not previously specified, subset of qubits, which might be different for different output branches, the situation changes. As shown in Ref. \cite{Fo06}, in this case maximally entangled pairs can be created from the  W state quasi-deterministically and one finds that above criteria can no longer be applied. This is true for any ``local'' measure (such as the localizable entanglement), while ``global'' measures (e.g., Entropic entanglement width, Schmidt rank width, Schmidt number, Geometric measure of entanglement) can still be used. \hfill $\diamond$ \end{rem}

\subsection{Geometric measure of entanglement}\label{sect_I_gm}

As a third measure we consider the \emph{geometric measure}
of entanglement $E_g$, a multipartite entanglement (type I)
monotone introduced in Ref. \cite{Sh95}, and show that also
$E_g$ can be used to obtain a criterion for  universality.
This measure is defined as follows. Let $|\psi\rangle$  be
an $N$-qubit state, and let $\pi(|\psi\rangle)$ denote the
maximal modulus squared of the overlap between
$|\psi\rangle$ and a complete product state on $N$ qubits,
\be
\pi(|\psi\rangle)=\max_{|\varphi\rangle=|\varphi_1\rangle\otimes
\ldots \otimes|\varphi_N\rangle}
|\langle\psi|\varphi\rangle|^2. \ee Then the geometric
measure is defined by \be E_{g}(|\psi\rangle):= - \log_2
\pi(|\psi\rangle).\ee It was proven in Ref. \cite{Sh95}
that $E_g$ is a type I entanglement monotone.  Moreover,
this measure trivially satisfies property (ii) stated in
theorem \ref{thm_monotone}, i.e., one has
$E_g(|\psi\rangle|0\rangle) = E_{g}(|\psi\rangle)$.
Therefore, one finds that $E_g$ is also a type II monotone.
In order to compute the supremum $E_g^*$, we use that the
geometric measure of the $d\times d$ cluster states grows
as $O(d^2)$ \cite{Ma06}, which implies that $E_g^*=\infty$.
This leads to the following criterion.

\begin{thm}\label{thm_ED_gm}
Let $\Psi$ be a universal resource.  Then $E_g(\Psi)=
\infty$.
\end{thm}
Thus, any resource with a bounded geometric measure, cannot
be universal.

{\it Example.---} The geometric measure of the  $N$-qubit W
state is equal to \cite{We03}   \be
E_g(|W_N\rangle)=(N-1)\log_2 \frac{N}{N-1},\ee which, for
large $N$, tends to $(\ln 2)^{-1}\sim 1.44$. This shows
that $E_g$ is bounded for the $W$-states, such that these
states can not form a universal resource.\hfill $\diamond$

\subsection{Schmidt measure}\label{sect_I_sm}

Finally, as a fourth measure we consider the Schmidt
measure, a (type I) entanglement monotone introduced in Ref. \cite{Ei01}.
The Schmidt measure $E_s(|\psi\rangle)$ of an $N$-qubit
state $|\psi\rangle$ is defined as the logarithm to base
two of the minimal number $K$ of $N$-qubit complete product
states $\{|\alpha_1\rangle, \dots, |\alpha_K\rangle\}$,
such that $|\psi\rangle$ can be written as a linear
combination \be|\psi\rangle = \sum_{i=1}^K
a_i|\alpha_i\rangle,\ee for some complex coefficients
$a_i$. As it can easily be shown that
$E_s(|\psi\rangle|0\rangle) = E_s(|\psi\rangle)$ and as
$E_s$ is a genuine (type I) entanglement monotone, it
follows that the Schmidt measure is a type II monotone.
Moreover, one has calculated the Schmidt measure of the 2D
cluster states in Ref. \cite{He04}, yielding $E_s(|C_{d\times
d}\rangle)= O(d^2)$, which shows that $E_s^*=\infty$. We
can therefore conclude the following:
\begin{thm}\label{thm_ED_sm}
Let $\Psi$ be a universal resource.  Then
$E_s(\Psi)=\infty$.
\end{thm}
{\it Example.---} Consider the GHZ states \be|\psi_{\mbox{\scriptsize{GHZ}}}\rangle:= \frac{1}{\sqrt{2}}\left( |0\rangle^{\otimes n} + |1\rangle^{\otimes n} \right).\ee  By definition,
these states have a constant Schmidt measure (equal to
$\log_2(2) = 1$), showing that also by this criterion (see also section \ref{sect_I_ewd}) these
states cannot yield universal resource.  \hfill
$\diamond$

\section{Efficient universality and scaling of entanglement}\label{sect_I_eff}

In order to obtain criteria for universality, so far we
have not considered \emph{efficient} universality, as
defined in section \ref{subsect_def_eff}. Evidently, every
efficient universal resource must also satisfy theorem
\ref{thm_P1}, and, conversely, any resource which is
identified by corollary \ref{cor_criterion} as not being a
universal resource, a fortiori cannot be efficiently
universal. However, when efficient universality is taken
into account, the criterion in corollary
\ref{cor_criterion} can considerably be strengthened, as we
will show in this section.

\subsection{General strategy}

Our approach will be illustrated by the following example,
where we focus on the entropic entanglement width. Consider
an efficient universal resource \be \Psi:=
\{|\psi_1\rangle, |\psi_2\rangle, \dots\}, \ee where
$|\psi_i\rangle$ is a state on $N_i$ qubits, for every
$i=1,2,\dots$. For every $d\times d$ cluster state on
$n=d^2$ qubits, there exists a state
$|\psi_{f(n)}\rangle\in\Psi$ in the resource $\Psi$ (where
$f(n)\in\{1,2,\dots\}$), such that
$|\psi_{f(n)}\rangle\geq_{\mbox{\tiny LOCC}}|C_{d\times
d}\rangle$, thus showing that \be E_{\mbox{\scriptsize
wd}}(|\psi_{f(n)}\rangle)\geq E_{\mbox{\scriptsize
wd}}(|C_{d\times d}\rangle) \geq O(\sqrt{n}).\ee In the
last equality we have used that the rank width of the $(2l
+1)\times (2l+1)$ grid graph is larger then $l-1$, as
proven by Oum \cite{Oum_private}. This implies that the
entropic entanglement width of the $d\times d$ cluster
states scale (at least) as $O(d) =O(\sqrt{n})$. As $\Psi$
is an efficient universal resource,  the number of qubits
$N_{f(n)}$ on which $|\psi_{f(n)}\rangle$ is defined, only
grows polynomially with the size $n$ of the 2D cluster
states (see observation \ref{obs2} in section
\ref{subsect_obs2}). Thus, there exists a polynomial $p(n)$
such that $N_{f(n)}\leq p(n)$. Conversely, for large $n$
(where we only consider the leading order, denoted by $k$,
of the polynomial $p(n)$), one has $n\geq
O((N_{f(n)})^{1/k})$, thus showing that
\be\label{scaling_width} E_{\mbox{\scriptsize
wd}}(|\psi_{f(n)}\rangle)\geq O((N_{f(n)})^{1/(2k)}).\ee We
therefore find a necessary condition on the \emph{scaling
behavior} of the entanglement measure $E_{\mbox{\scriptsize
wd}}$ on any efficient universal resource $\Psi$: equation
(\ref{scaling_width}) shows that $E_{\mbox{\scriptsize
wd}}(|\psi_i\rangle)$ must grow as $O((N_i)^{1/(2k)})$ for
some $k$, which is essentially equivalent to stating that
this measure must grow faster-than-logarithmically with the
system size on $\Psi$. Hence, resources where the entropic
entanglement width grows at most as the logarithm of the
system size, cannot be efficient universal resources  (in
particular, resources with bounded entanglement width are
covered by this result).

The argument of the above example can be repeated for all
type II monotones $E$. In every case, \emph{a
necessary condition on the scaling behavior of $E$ for
efficient universal resources  is obtained}. More
formally, one has the following.

\begin{thm}\label{thm_scaling}
Let $\Psi:= \{|\psi_1\rangle, |\psi_2\rangle, \dots\},$ be
a resource, where $|\psi_i\rangle$ is a state on $N_i$
qubits, for every $i$. Let $E$ be a type II monotone, and
let $\varphi$ be a function such that, for every
\footnote{Here, the words 'for every 2D cluster state' can be replaced by
'for infinitely many 2D cluster states' in order to yield a slightly stronger
result.} 2D cluster state $|C_{d\times d}\rangle$ on
$n=d^2$ qubits, one has \be E(|C_{d\times d}\rangle) \geq
\varphi(n).\ee If $E(|\psi_i\rangle)$ scales as $\log
\varphi(N_i)$, then $\Psi$ cannot be an efficient universal
resource.
\end{thm}

This result should be regarded as a strengthening of
theorem \ref{thm_P1}. We emphasize again that the 2D cluster
states in principle do not play a distinguished role in
theorem \ref{thm_scaling}, in the sense that they can be
replaced---without weakening or strengthening the
result---by arbitrary efficient universal resources  or, in fact, any family of states
\emph{which themselves also can efficiently be prepared.}
However, as pointed out before, being graph states, the 2D
cluster states usually form a suitable test-bed also in
this case.

Specializing these considerations to entanglement width, localizable
entanglement, geometric measure and Schmidt measure, one
finds:

\begin{thm}
Let $\Psi$ be a resource. If one of the measures
$E_{\mbox{\scriptsize wd}}$, $\chi_{\mbox{\scriptsize
wd}}$, ${\cal N}_{\mbox{\scriptsize{LE}}}$, $E_g$ or $E_s$
scales at most logarithmically with the system size on
$\Psi$, then $\Psi$ cannot be an efficient universal
resource.
\end{thm}
The proof of this result simply follows from the fact that
all these measures scale as $O(n^{1/k})$, for some $k$, for
the 2D cluster states on $n$ qubits.

When considering entanglement width, the above criterion allows one to determine new examples of
no-go results which were not detected by theorem
\ref{thm_P1}.

{\it Example.---} Consider, as at the end of section
\ref{sect_criterion_width}, a family of ground states
$\{|\phi_N\rangle\}$ of a 1D spin system, where
$|\phi_N\rangle$ is a state of $N$ spins. Investigating now
the \emph{critical} phase, one typically has that the block
entanglement entropy of a subchain of spins of length
$L\leq N$ scales as $\log L$ \cite{Vi03}. Similar to the
example in  section \ref{sect_criterion_width}, one then
finds that the entanglement width of such a ground state
can at most scale as the logarithm of the system size. This
shows that\emph{ any resource of ground states of 1D
(critical) spin systems where the block entanglement of a
subchain scales as the logarithm of the length of the
subchain, cannot be an efficient universal resource.} Thus,
together with the example in section
\ref{sect_criterion_width}, we find that ground states of
critical or non-critical 1D spin systems can typically not
be efficient universal resources. \hfill $\diamond$

Another interesting no-go result is obtained using the Schmidt measure.

{\it Example.---} One easily finds that the Schmidt measure
can grow at most logarithmically with the system size for
the W states (this immediately follows from the definition
of these states). This  implies that the W states cannot be
efficient universal resources.  In contrast to the argument
based on localizable entanglement (see also section
\ref{sect_I_le}), the argument presented here is also valid
when considering the possibility of random output
particles. Note also that by considering the geometric
measure, as in section \ref{sect_I_gm}, one can conclude
that the W states do not form an (efficient) universal
resource. \hfill $\diamond$

\subsection{Efficient universality and classical simulation of MQC}\label{sect_I_sim}
It is instructive to point out a relation between efficient
(exact deterministic)  universality of a resource $\Psi$ and
the problem of efficiently (i.e.,  with polynomial
overhead) classically simulating MQC on this resource. When a resource is identified as \emph{not} being an efficient universal resource, one may wonder whether an efficient classical simulation of MQC on this resource becomes possible. Although  one can of course not expect that such a result would hold in general,  here we note that such a property is valid if one considers the criteria for efficient universality corresponding to the Schmidt-rank width and the Schmidt measure.  In particular, in previous work \cite{Va06a} we showed that
MQC can be simulated on every resource where the
Schmidt-rank width grows only logarithmically with the
system size. This shows that any resource which is ruled
out by the Schmidt-rank width criterion in theorem
\ref{thm_scaling} as an efficient universal resource, allows an efficient classical
simulation of MQC.

A similar remark can be made regarding the Schmidt measure.
Suppose that $\Psi$ is a resource where the Schmidt measure
grows at most logarithmically with the system size.
Equivalently, every state in the family $\Psi$ can be
written as a linear combination of polynomially many
(w.r.t. the number of qubits) complete product states. One
then easily verifies that any LOCC on such states can be
simulated with polynomial overhead by a classical computer.
In  particular, the expectation value of a local observable
$O:=O_1\otimes\dots\otimes O_n$ can be efficiently computed
when an $n$-qubit system is in a state $|\psi\rangle$ with
logarithmically scaling Schmidt rank: writing
\be|\psi\rangle = \sum_{i=1}^K a_i|\alpha_i\rangle,\ee
where the $\{|\alpha_i\rangle\}$ are $K=$ poly$(n)$
complete product states, the expectation value \be \langle
\psi| O|\psi\rangle = \sum_{i, j=1}^K a_i\bar a_j \langle
\alpha_i| O|\alpha_j\rangle\ee can efficiently be computed,
as every term $\langle \alpha_i| O|\alpha_j\rangle$ can be
efficiently computed and as there are only poly$(n)$ such
terms in the above sum. Similarly, a state obtained after
performing a local projection on $|\psi\rangle$ can
efficiently be determined (note that such states can have a
Schmidt measure of at most $\log_2 K$). One can therefore
conclude that any resource which fails to pass the Schmidt
measure criterion in theorem \ref{thm_scaling}, even allows
an efficient classical simulation of MQC.

Thus, it is interesting to find that in the issue  of
efficient universality, as well as in the context of
efficient simulation of MQC,  the \emph{scaling} of
entanglement with the system size plays an important role.

\section{Examples of efficient universal resources}\label{go}

To provide examples of universal resource states, our
strategy is to establish an equivalence, up to a polynomial
overhead, among (efficient) universal resources. The proof
of efficient universality of the 2D cluster states
\cite{Ra01, Ra03} means an efficient capability to simulate
any circuit model by LOCC and, in particular, to prepare
any other resource state by LOCC. Thus, all we have to show
for other resource states is that the opposite
transformation is also efficiently possible.

\begin{thm}
The graph states corresponding to the 2D square, hexagonal, triangular and
Kagome lattices are efficient universal resources for MQC.
\end{thm}

\begin{figure}[b]
\hspace{2.2cm}\begin{minipage}{14cm} 
\includegraphics[width=5.5cm,clip]{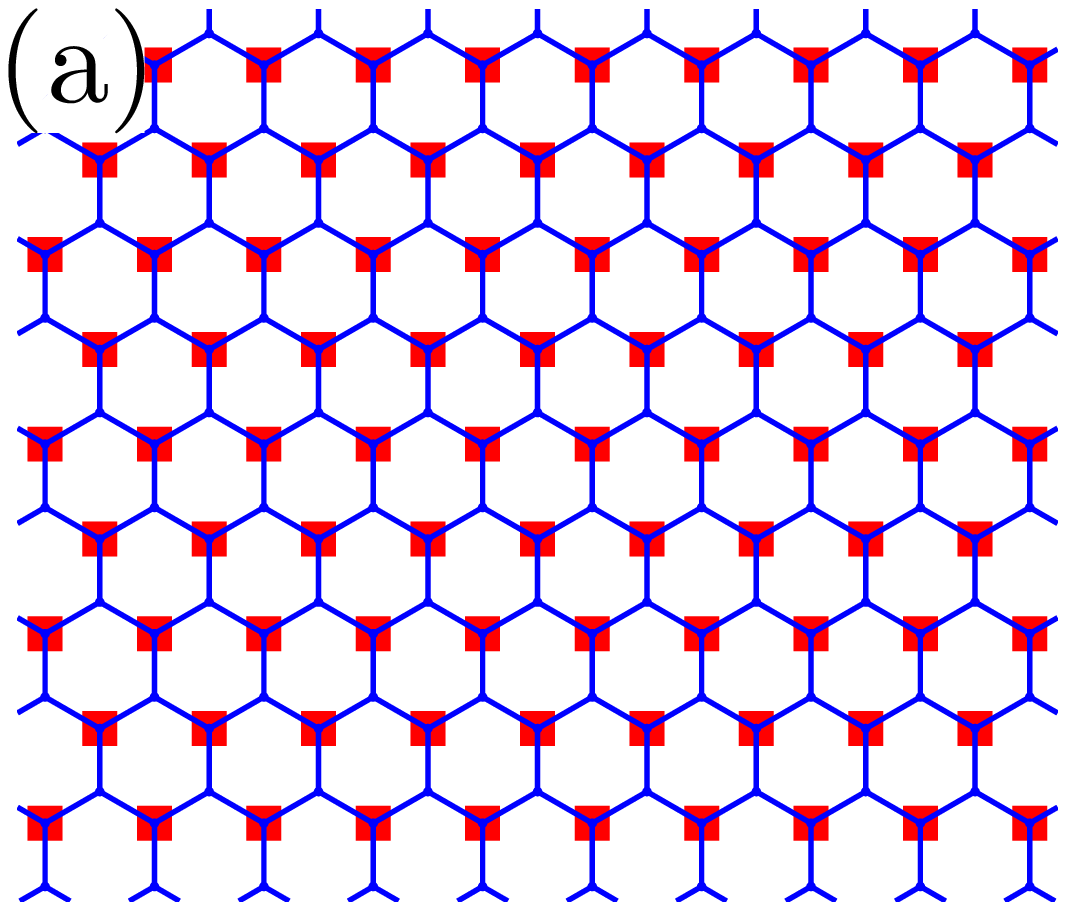}
\includegraphics[width=5.5cm,clip]{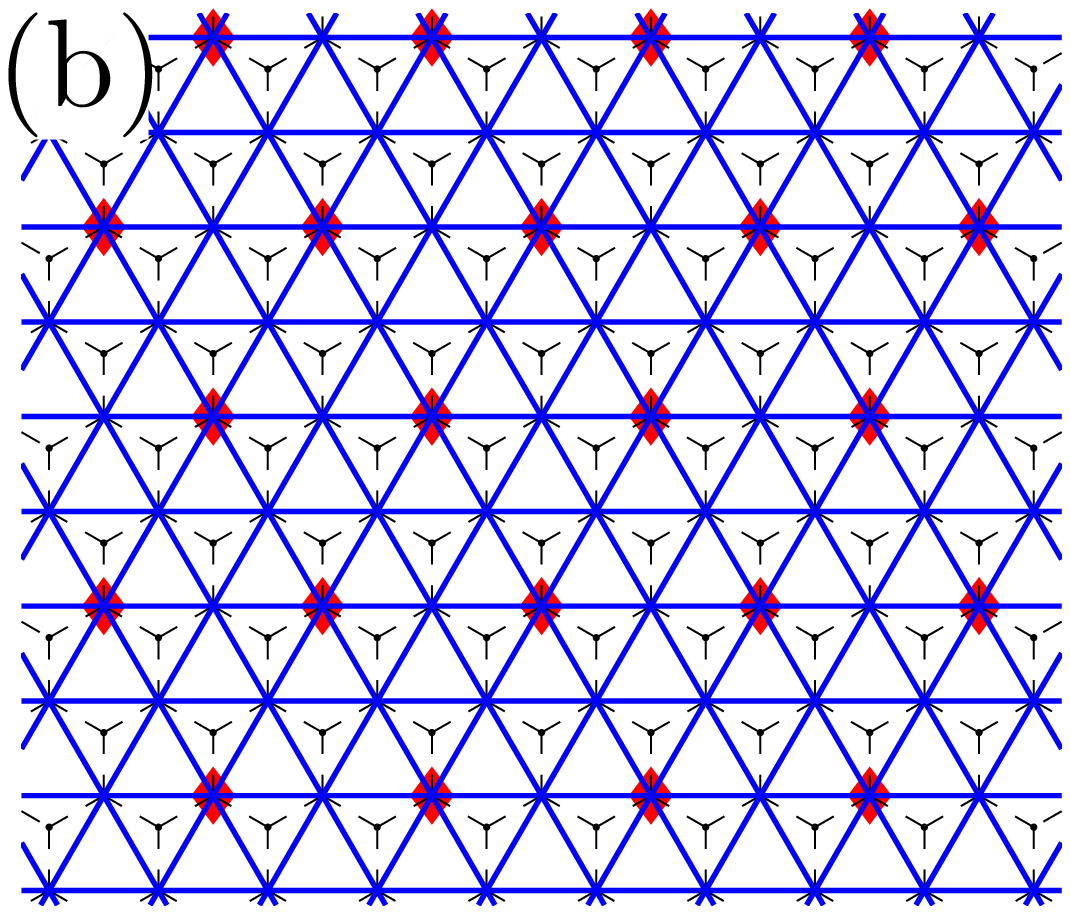}
\end{minipage}
\begin{minipage}{14cm}
\hspace{2.2cm}\includegraphics[width=5.5cm,clip]{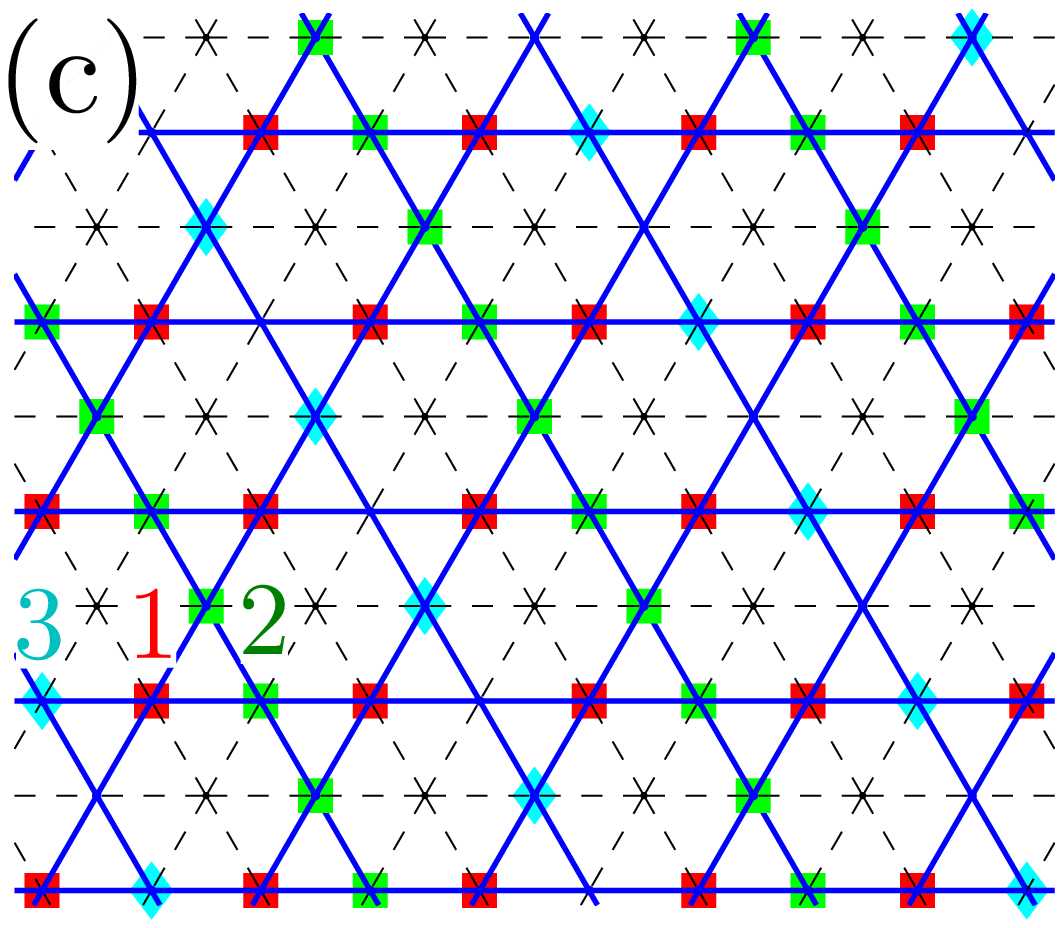}
\includegraphics[width=5.5cm,clip]{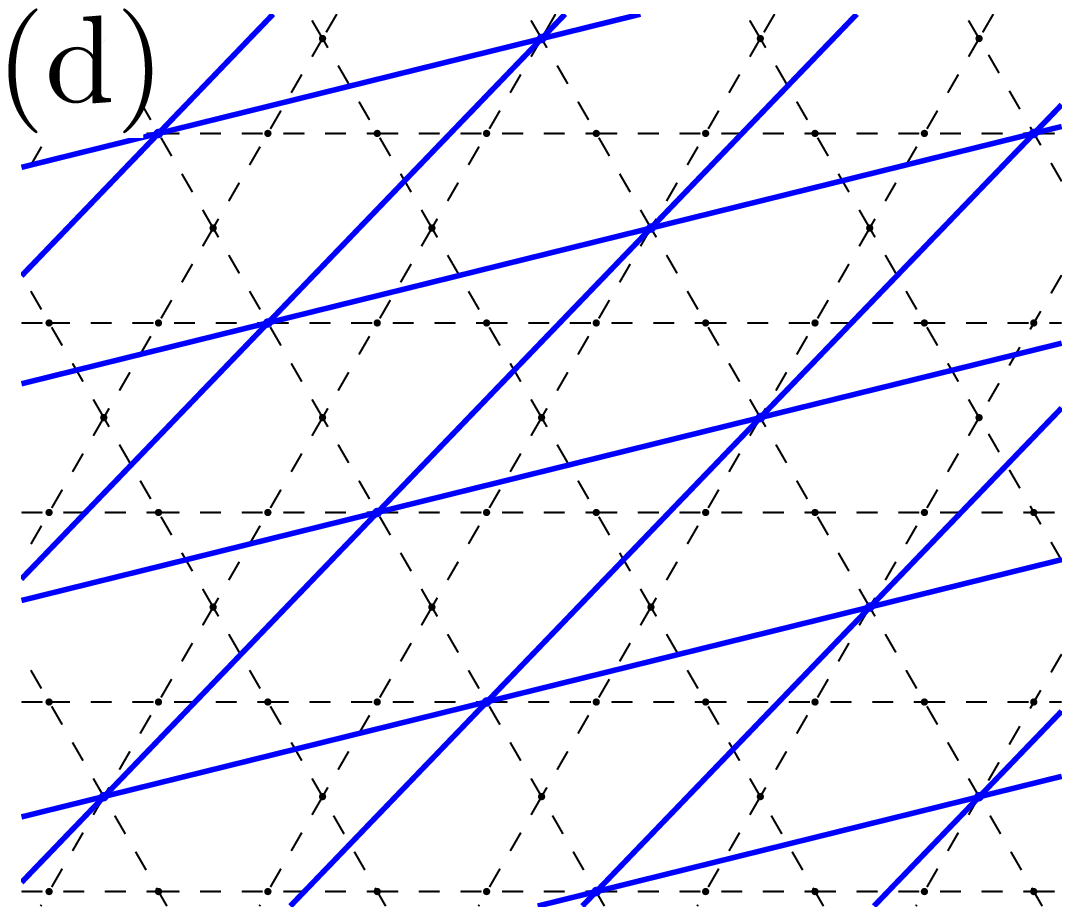}
\end{minipage}
\caption{Examples of efficient universal resource for MQC. These are
graph states corresponding to
(a) hexagonal, (b) triangular and (c) Kagome lattices.
Deterministic LOCC transformation from (a) to  (d) (2D cluster state)
via (b) and (c) is indicated, where simple graph rules can be used
sequentially ($\sigma_y$ and $\sigma_z$ measurements are displayed by
$\square$ and $\diamondsuit$, respectively).
The spatial overhead for the transformation is constant.}
\label{UniversalLattices}
\end{figure}

{\it Proof:}
According to our strategy, a proof is given in the form of an explicit
deterministic LOCC protocol which transforms such other candidate states
into the 2D cluster state with a polynomial overhead in resources and
operations (see Fig.~\ref{UniversalLattices}).

As LOCC, we translationally apply $\sigma_y$ and $\sigma_z$ measurements,
which are known to correspond to simple update rules of the graph,
namely local complementation (inversion of the neighborhood subgraph) and vertex deletion, respectively \cite{He06}.
We will get a resulting graph state {\it deterministically} after
applying local Clifford unitary corrections $u_{y,k}$ and $u_{z,k}$
on neighboring qubits (with help of classical communication), such as
\begin{eqnarray*}
u_{y,k} &=& \sqrt{(-1)^k (-i\sigma_z)} =
\frac{1}{\sqrt{2}} \left({\bf 1} - (-1)^k i\sigma_z\right) , \\
u_{z,k} &=& \sigma_z^k ,
\end{eqnarray*}
which depend on the measurement outcomes $k=\{0,1\}$.

Let us illustrate such graph transforms to prove the
theorem. Our LOCC protocol transforms (a) hexagonal-lattice
graph state into (d) the 2D cluster state via (b) the
triangular-lattice and (c) the Kagome-lattice graph state
sequentially. When we apply the $\sigma_y$ measurements on
vertices of the hexagonal lattice in
Fig.~\ref{UniversalLattices} ~(a), we will have a triangle
around the measured qubits by the local complementation.
The resulting graph turns out to be a triangular lattice
(b). Next, we apply the $\sigma_z$ measurements on vertices
marked by $\diamond$, which simply deletes the measured
qubits along with the attached edges. The graph now becomes
a so-called Kagome lattice, which has degree 4 uniformly in
a similar manner as the square lattice. The last step is a
bit involved, but  straightforward by applying $\sigma_y$
and $\sigma_z$ measurements according to the ordering of
Fig.~\ref{UniversalLattices}. Accordingly, we finally
obtain a square lattice, i.e., a 2D cluster state. The
total spatial overhead is eight in that a unit square on
(d) is obtained from eight hexagons in (a). Since the
overhead is constant, all resource states are efficient
universal. \finpr

Note that, since these other resource states
can be transformed into the 2D cluster state always by {\it adaptive local
projective measurements} assisted with classical communication, and local (Clifford) operations, they are
also universal in the sense of a conventional one-way quantum computation.

It is interesting to observe that the square lattice, together with the
hexagonal and triangular lattices are the only possibilities
to obtain a regular tiling of the 2D plane. Furthermore, the Kagome lattice
is an example of a uniform semiregular 2D tiling with 2 basic tiles
(the triangle and the hexagon). The hexagonal lattice has vertex degree 3,
which leads to an increased robustness against local noise as compared to
the 2D cluster state \cite{Du04}.
Since the 1D cluster state (with uniform vertex degree 2) has been proved not
to be universal \cite{Va06}, the vertex degree 3 is minimal for
universal resource on {\it uniform} lattice structures.

We remark that other universal resource states have been presented \cite{Ch05}
based on non-uniform lattice structures (see also \cite{Ta06}), where
each gate in a universal set of unitary gates can be implemented by
local measurements on an elementary unit and these units are combined
(bottom-up approach).
Here we took, in contrast, a top--down approach, where we prove universality
by explicitly constructing LOCC protocols that yield the 2D cluster state.

\begin{figure}[t]
\hspace{2.7cm}\includegraphics[width=10cm,clip]{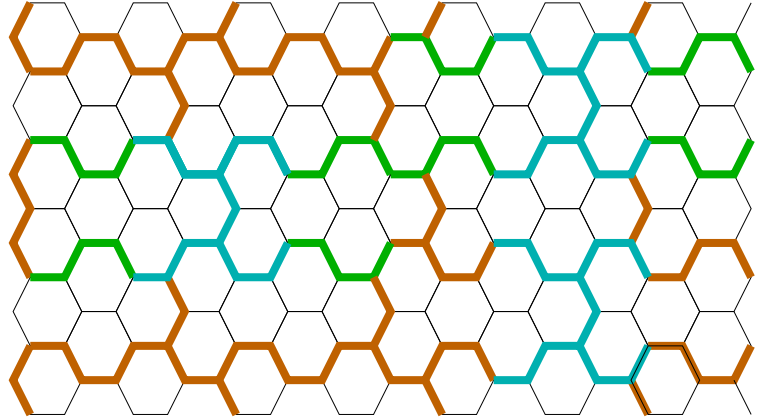}
\caption{A direct simulation of the circuit model on the
hexagonal lattice. After measuring uncolored qubits in the
$\sigma_z$ direction, the resource state is capable of
simulating a set of universal gates between any neighboring
two quantum wires (snaking horizontal paths) and to compose
them into an arbitrary unitary operation. Here we
illustrate an implementation (with the same measurement
pattern of Ref.~\cite{Ra01, Ra03}) of CNOT and of arbitrary
$SU(2)$ rotation with the Euler decomposition by the
15-qubit 90-degree rotated H part (blue) and the 5-qubit
chain part (green), respectively.} \label{fig:circuit_hex}
\end{figure}

Of course, we can simulate {\it directly} any circuit model
on such new universal resources in practice, without
transforming them into the 2D cluster state of the 2D
square lattice. Let us illustrate this by a hexagonal
lattice (due to its potential advantage for the local
decoherence compared with a square lattice). In
Fig.~\ref{fig:circuit_hex}, we show a direct simulation of
the circuit model on the 2D hexagonal lattice, after
deleting uncolored qubits by the $\sigma_z$ measurements.
We consider each snaking horizontal path running from the
left to the right as a quantum wire which corresponds to an
evolution of a logical single qubit in the circuit model.

Here we give an alternative proof of its efficient exact
deterministic universality in a similar manner as the proof
of the 2D cluster state \cite{Ra01, Ra03} (see also
Ref.~\cite{Ch05} for other resources). Namely, we take a
bottom-up approach by showing two (sufficient) conditions
are fulfilled for our family of resource states:
\begin{itemize}
\item[(i)]
{\it Simulation of a universal set of elementary gates}:
It is capable of implementing a set of universal gates on logical qubits
with at most polynomial overhead.
\item[(ii)]
{\it Composability}:
The simulation (i) of elementary gates is composable between any neighboring
pair of logical qubits, as well as repeatable.
\end{itemize}
Here we follow the implementation (measurement pattern) of
Ref.~\cite{Ra01, Ra03} of CNOT and of arbitrary $SU(2)$
rotation with the Euler decomposition in terms of the
15-qubit 90-degree rotated H part (blue) and the 5-qubit
chain part (green), respectively. The capability of
implementing such gates as often as required follows
immediately from its periodic structure. The length of an
extra one-dimensional path can be adjusted, if necessary,
by the $\sigma_{y}$ measurement.

A slightly more economical simulation can  also be done by
selecting the controlled-phase gate and the set $SU(2)$ of
single-qubit rotations as a set of elementary gates for the
circuit model. By measuring the middle qubit of the
``bridge between quantum wires'' (i.e., the center qubit in
the blue part) in the $\sigma_y$ direction, we can simply
implement a logical ``remote'' controlled-phase gate (cf.
Ref.~\cite{Ch05}), without additional measurements along
the quantum wires.

\section{Encoded universality}\label{encoded}

So far we have restricted our attention to  situations
where the resource states consist of systems with fixed
dimension $d$, and where the desired target states are
multipartite states of $d$-dimensional systems. We have
concentrated on $d=2$, i.e., qubit systems, and have
provided definitions and criteria when such a family of
resource states is capable of producing any other quantum
state of $n$ qubits. In this section we will show how to
extend our results in order to take higher-dimensional
systems, generation of systems with different dimensions,
and {\em encoded} quantum states into account. We will
first specify what we mean by encoding, and then put
forward a definition for encoded (CQ-)universality in MQC,
based on different notions of {\em locality}. We will
finally discuss the applicability of entanglement criteria
developed in the previous sections, and discuss examples of
encoded universal resources.

\subsection{Encoded quantum states}\label{encoding}

We start by some general considerations on encoded quantum states.

\subsubsection{Qubits and qudits}\label{dlevel}

In the measurement-based model for quantum computation, a
resource state constituting of $N$ two-dimensional systems
is processed by sequences of LOCC. It is clear that the
physical systems remain the same throughout the procedure,
and hence the resulting quantum state is also one
consisting of qubits. It is {\em impossible} to generate a
state of $d$-dimensional systems (qudits) from a qubit
system. More generally, if the system constituting the
resource state are $k$-dimensional, and the states to be
generated constitute of $d$-dimensional systems, then one
cannot generate any such state whenever $k \not= d$. What
is, however, possible, is to generate the desired state in
an {\em encoded} form. In  almost all conceivable cases,
the generation of a quantum state in such an encoded form
is sufficient.

The most natural kinds of encoding are the following:
\begin{itemize}
\item[(i)] If $k \geq d$, then ideally one encodes $m=\lfloor\log(k)/\log(d)\rfloor$ $d$-level systems into one $k$--level system.
\item[(ii)] If $k < d$, then ideally one uses $m=\lceil\log(d)/\log(k)\rceil$ $k$-level system to encode one $d$-level system.
\end{itemize}
In neither of these two cases additional levels are used.
One may also consider more complex encodings, e.g., any
encoding that can be generated by a poly-sized quantum
circuit. However, in this case quantum information
corresponding to a $d$-level system may be distributed
among {\em all} $k$-level systems. In  the following, we
will restrict ourselves to the kinds of encodings specified
above, where a natural association with physical systems is
maintained. This restriction allows us to use
entanglement-based criteria for universality also for
encoded systems.

\subsubsection{General encodings}

The treatment of systems of different dimensionality is not
the only case where encodings are useful. Even if one is
restricted to qubits, it is natural to consider the
question whether states can be generated in an encoded way.
Encoded states appear naturally in, e.g.,  quantum
computation once error-correction is considered. In
particular, any fault-tolerant implementation of quantum
computation actually generates states in an encoded form,
and performs unitary operations on the encoded states. In
this case the (redundant) encoding serves to protect
quantum information against the influence of errors. One
may consider encodings also for different reasons. For
instance, for certain physical set-ups it may be impossible
(or very hard) to perform arbitrary single-qubit operations
and two-qubit gates. However, for certain encodings the
implementation of logical single- and two-qubit gates,
i.e., gates acting on the encoded or logical systems, is
possible. Such a situation occurs for instance in
charge-controlled quantum dots, where despite of
difficulties to obtain inhomogeneous single-spin rotations,
processing of encoded quantum information is possible
\cite{DiVi00, Ta05, Ta05b}. Finally---and this brings us
back to the present study of universality---if one is
ultimately interested in a CC computational scheme, it is
often sufficient that a quantum computer is capable of
generating states in an encoded form. In particular, we
will see below that any encoded CQ-universal resource for
MQC is also CC-universal.

For simplicity, in the following  we will restrict
ourselves to the case where $k=d=2$. Generalization to $k
\not= d \geq 2$ is, however, straightforward.

Next we define what  is meant by ``encoded qubits'' and
``encoded quantum states''. First we consider encodings of one-qubit systems and states.

\begin{defn}{\bf (Encoded qubit):} An $m$-qubit encoding ${\cal E}:= ({\cal H}, \{|{\bm 0}\rangle, |{\bm
1}\rangle \})$
of a one-qubit system is specified by
\begin{itemize}
\item a Hilbert space ${\cal H}\cong \mathbb{C}^{2^m}$, representing a system of $m$ qubits, and
\item two (fixed) orthogonal states $|\psi_0\rangle, |\psi_1\rangle\in {\cal H}$. We use the shorthand notation $|{\bm0}\rangle :=|\psi_0\rangle\mbox{ and } |{\bm
1}\rangle :=|\psi_1\rangle.$
\end{itemize}

Let $|\phi\rangle=a|0\rangle + b|1\rangle$ be an arbitrary one-qubit state. The encoded version  of $|\phi\rangle$ with respect to the encoding ${\cal E}$ is the state $|{\bm \phi}\rangle:= a|\bm 0\rangle + b|\bm 1\rangle\in{\cal H}$.
\end{defn}
This definition is now generalized to encompass encodings of many-qubit systems and states.

\begin{defn}{\bf (Encoded $n$-qubit system):} An $m$-qubit encoding ${\cal E}:=({\cal H}^E, {\cal P}, \{|{\bm 0}\rangle, |{\bm
1}\rangle \})$
of an $n$-qubit system is specified by
\begin{itemize}
\item a Hilbert space ${\cal H}^E\cong \mathbb{C}^{2^{mn}}$, representing a system of $m\cdot n$ qubits labeled by a set $E=\{1, \dots, m\cdot n\}$,
\item a partition ${\cal P}:=\{A_1, \dots, A_n\}$ of $E$, where every $A_k$ denotes a set of $m$ qubits
of $E$, and
\item two (fixed) orthogonal $m$-qubit states $\{|{\bm 0}\rangle, |{\bm
1}\rangle \}$.
\end{itemize}

Consider an $n$-qubit state \be |\phi\rangle =
\sum_{i_1,i_2, \ldots i_n\in\{0,1\}^n } c_{i_1 i_2 \ldots
i_n} |i_1\rangle|i_2\rangle\ldots |i_n\rangle.\ee The
encoded version of $|\phi\rangle$ w.r.t. the encoding
${\cal E}$ is the state \be |{\bm \phi}\rangle =
\sum_{i_1,i_2, \ldots i_n\in\{0,1\}^n } c_{i_1 i_2 \ldots
i_n} |{\bm i_1}\rangle_{A_1}|{\bm i_2}\rangle_{A_2} \ldots
|{\bm i_n}\rangle_{A_n}\in{\cal H}^E.\nonumber\ee Here the
subscripts $A_k$ denote on which qubits the states $|{\bm
i}_k\rangle$ are defined.
\end{defn}

For notational simplicity, in the definition above we have
considered the same encoding for every qubit.
Generalization to different encodings for different qubits
is straightforward.

Similarly, for $k \not= d \geq 2$  an encoded qudit can be
defined by a set of $d$ orthogonal states of $m$ $k$-level systems, and
the definition of an encoded state follows immediately. The
case of generating $d$-level systems using $k$-level
systems is also covered, where the encoding corresponds to
(i) or (ii) specified in Sec. \ref{dlevel}.

\subsection{Local operations on encoded systems}

The notion of {\em locality} is central in MQC. In
particular, entangled quantum states are processed by local
operations and classical communication. This allows one to
attribute entanglement as a resource to the initial state
of the system, and only in this context criteria based on
entanglement properties of resource states as presented in
section \ref{sect_ED_criteria} become meaningful. When
considering systems with a fixed dimension, e.g., qubits,
as we did in the previous sections, the notion of locality
is unique and simply corresponds to single-qubit
operations. When considering encoded systems as we do here,
one may, however, consider two different notions of
locality, and the corresponding classes of ``local''
operations assisted by classical communication.
\begin{itemize}
\item[(i)] Local with respect to physical systems
constituting the resource state: LOCC \item[(ii)] Local
with respect to the encoded systems: {\bf L}OCC
\end{itemize}
It depends on the context at hand  which of these two
notions of locality is of interest. The first notion,
associating locality to the physical systems constituting
the quantum state, seems to be the most natural one.
However, in some cases one might also naturally consider
(ii), i.e., locality with respect to logical or encoded
system. Notice that in the
case of {\bf L}OCC, joint operations on certain groups of
physical qubits are allowed. Hence, if we consider LOCC in
the definition for universality, we obtain a more stringent
requirement as when using {\bf L}OCC. Usage of {\bf L}OCC
will turn out to be useful when considering
entanglement-based criteria for encoded universality.

\subsection{Definition of encoded universality}
In this section, we precisely define what is meant by an
``encoded universal resource for MQC''. In order to do so,
we first need to specify the notion of a ``consistent
family of encodings'', which is the following. In the
present study, the role of  an infinitely large resource is
played by a family of states $\Psi=\{|\psi_1\rangle,
|\psi_2, \rangle, \dots\}$, where $|\psi_i\rangle$ is a
state on $N_i$ qubits and where $N_{i}\leq N_{i+1}$ for
every $i$. More precisely, the state $|\psi_i\rangle$
belongs to a Hilbert space ${\cal H}^{E_i}$, where $E_i$
denotes a set of $N_i$ qubits, such that $E_i\subseteq
E_{i+1}$, for every $i$. We will say that $\{{\cal
H}^{E_1}, {\cal H}^{E_2}, \dots\}$ is a \emph{nested}
family of Hilbert spaces. Now, suppose that, for every
Hilbert space ${\cal H}^{E_i}$  an encoding ${\cal E}_i =
\{{\cal H}^{E_i}, {\cal P}_i, \{|\bm 0\rangle,
|\bm1\rangle\}\}$ is given. Since a resource $\Psi$ plays
the role of an infinitely large state, when considering
encodings---together with the corresponding partitions of
the systems ${\cal H}^{E_i}$ into logical qubits as
specified by the encodings ${\cal E}_i$---it is natural to
impose a certain degree of consistency of these encodings.
In particular, we will require that the way in which the
Hilbert space ${\cal H}^{E_i}$ is partitioned into logical
qubits, should be consistent with the way in which ${\cal
H}^{E_{i+1}}$ is partitioned, i.e., ${\cal
P}_i\subseteq{\cal P}_{i+1}$ for every $i$. In such case,
we will call ${\cal F}:=\{{\cal E}_1, {\cal E}_2, \dots\}$
a \emph{consistent family of encodings}.

We are now in a position to give a definition of encoded
universal resource for MQC.

\begin{defn}\label{def_encoded_UR}{\em \bf (Encoded universality):}
Consider a nested family of Hilbert spaces $\{{\cal
H}^{E_i}\}_i$, and let $\Psi=\{|\psi_1\rangle, |\psi_2
\rangle, \dots\}$ be a family of states such that
$|\psi_i\rangle\in {\cal H}^{E_i}$, for every $i$. Let
${\cal F}:=\{{\cal E}_1, {\cal E}_2, \dots\}$ be a
consistent family of $m$-qubit encodings of this family of
Hilbert spaces. The family $\Psi$ is called an encoded
universal resource for MQC with respect to ${\cal F}$, if
for every $n$ and for every $n$-qubit quantum state
$|\phi_{\rm out}\rangle$, there exists a state
$|\psi_i\rangle \in \Psi$ such that $|\psi_i\rangle
\geq_{\mbox{\tiny LOCC}} |\bm \phi_{out}\rangle$, where
$|{\bm \phi}_{out}\rangle$ is the encoded version of
$|\phi_{out}\rangle$ w.r.t. the encoding ${\cal E}_i$.

The family $\Psi$ is said to be an encoded universal
resource for MQC if there exist a consistent family of encodings with respect to
which $\Psi$ is universal.
\end{defn}

\begin{rem} {\it General comments regarding definition \ref{def_encoded_UR}.---}

 (i) Any resource that is universal is also
encoded universal, although the opposite is not necessarily
true. We will discuss examples for such cases later in this
section.

(ii) Although we allow for an arbitrary one-qubit encoding,
the encoding needs to be fixed, i.e., it needs to be the
same for all output states.

(iii) As in the case of non-encoded universality, we also
assume here that the output state is prepared on the same
output particles in all branches of the LOCC protocol (an
extension to random output particles is possible in the
same way as in the non-encoded case).

(iv) In definition \ref{def_encoded_UR} we refer to LOCC,
i.e., operations on the individual physical systems (see
(i) in the preceding section). One may also consider a
weaker form of universality with respect to {\bf L}OCC.

(v) In the above definition, the same encoding is used for
every qubit. This restriction can easily be lifted without
changing the results to be presented in the following. We
have adopted this version of the definition for notational
convenience.

(vi) In this paper we do \emph{not} consider general
encodings by means of arbitrary poly-sized quantum
circuits. This is a restriction in the sense that the
latter leads to a more general definition of encoded
universality (i.e., it is likely that there exist encoded
universal resources w.r.t. such general codings which are
not covered by the present theory). The reason for this is
that it is not clear how criteria stated in terms of
entanglement can be formulated for such general encodings,
whereas this will be possible for the current definition.
However, it is clear that general encodings by poly-sized
quantum circuits deserve further attention. \hfill
$\diamond$
\end{rem}

We will sometimes consider a slightly weakened version of
encoded universality, in the following sense. Let $\{|\bm
0\rangle, |\bm 1\rangle\}$ be an ($m$-qubit) encoding of a
one-qubit system, let  $|\bm \phi\rangle$ be an encoded
version  of an $n$-qubit state $|\phi\rangle$ with respect
to this encoding (see definition \ref{def_encoded_UR}), and
let $U=U_1\otimes\dots\otimes U_n$ be an $n$-qubit local
operation, where \be U_k:= a |0\rangle\langle 0| +
b|0\rangle\langle 1|+c|1\rangle\langle 0|+d|1\rangle\langle
1|\ee (for every $k=1, \dots, n$). Then the encoded version
${\bm U }={\bm U}_1\otimes\dots\otimes {\bm U}_n$ of this
operator is defined by \be {\bm U}_k:= a|\bm
0\rangle\langle \bm 0| + b|\bm 0\rangle\langle \bm 1| + c
|\bm 1\rangle\langle \bm 0|+ d|\bm 1\rangle\langle \bm
1|,\ee and naturally acts as \be {\bm U}|\bm \phi\rangle=
\sum_{{ i_1,i_2, \ldots i_n =0}}^{1} c_{i_1 i_2 \ldots i_n}
{\bm U}_1|{\bm i_1}\rangle_{A_1}  \ldots {\bm U}_n|{\bm
i_n}\rangle_{A_n}.\nonumber\ee For example, given an
encoding $\{|\bm 0\rangle, |\bm 1\rangle\}$ the logical
single-system Pauli operations are defined by \be
{\bm \sigma_z} = |\bm 0\rangle\langle \bm 0| - |\bm 1\rangle\langle \bm 1|, \nonumber\\
{\bm \sigma_x} = |\bm 0\rangle\langle \bm 1| + |\bm 1\rangle\langle \bm 0|, \\
{\bm \sigma_y} = -i|\bm 0\rangle\langle \bm 1| + i|\bm 1\rangle\langle \bm 0|.\nonumber
\ee Note that an operation which is local in the sense of logical qubits need not be local at the level of the physical qubits in the system, i.e., it may act jointly on several physical qubits.

We will sometimes consider resources which are encoded universal in the restricted sense that it is possible to prepare an encoded version of any state \emph{up to logical local operations}. Notice that a resource which is universal in this (restricted) sense with respect to CQ is, however, fully CC-universal. That is, if we are only interested in obtaining classical outputs, such logical local operations do not play a role as we simply have to adapt the basis of the final measurements. We also note that the 2D cluster states are CQ-universal under projective local measurements up to Pauli operations; this property, together with the above restricted notion of encoded universality,  will play a role in the next section.

\subsection{Encoded universality and the 2D cluster states}
In this section, it is our aim to relate encoded universal
resources to (encoded versions of) the ``standard''
universal resource, the family of 2D cluster states, i.e.,
this section will contain a result analogous to observation
\ref{obs1} in section \ref{sect_obs}. Although this result
will be highly analogous to observation \ref{obs1}, its
proof will not be as straightforward and will require some
extra work. The result is the following.

\begin{thm}\label{obs6}
A family of states $\Psi$ is  an encoded universal resource
with respect to a specific encoding (more precisely: w.r.t.
a consistent family of encodings) up to logical Pauli
operators, if and only if encoded 2D cluster states $|\bm
C_{d \times d}\rangle$ of arbitrary size $d$ with respect
to the same encoding can be generated by means of LOCC, up
to logical Pauli operations.

A family of states $\Psi$ is an encoded universal  resource
up to logical Pauli operators if and only if there exists
an encoding (more precisely: a consistent family of
encodings) such that encoded 2D cluster states $|\bm C_{d
\times d}\rangle$ of arbitrary $d$ can be generated by
means of LOCC, up to logical Pauli operations.
\end{thm}
{\it Proof: } Necessity of the condition  is obvious, as an
encoded universal resource must be capable of producing an
encoded version of any state, in particular an encoded 2D
cluster state. Sufficiency of the condition is based on the
fact that an encoded 2D cluster state is an encoded
universal resource under LOCC up to logical Pauli
operations for {\em any} encoding. The proof of the latter
statement is postponed until Theorem
\ref{encodedEUR}.\finpr

Note that it is by no means trivial that  local operations
on physical qubits allow one to process encoded quantum
information. E.g., as already indicated above, logical
local operations need not act locally on the physical
qubits, i.e., one may not be able to realize logical local
operations as physical local operations. For example,
consider a 3-qubit encoding where $|\bm 0\rangle:=
|000\rangle$ and $|\bm 1\rangle:= |W_3\rangle$, where the
latter is the W state on 3 qubits. It is clear that in this
case a logical $\sigma_x$ operation, i.e., \be {\bm
\sigma}_x:= |000\rangle\langle W_3| + |W_3\rangle\langle
000| \ee acts non-locally on the physical qubits.

However, we will find that it is always possible to
realize logical projective two-outcome measurements (which
are the basic ingredient of the one-way 2D cluster state
model) by physical local operations. We again refer to
Theorem \ref{encodedEUR}.

\subsection{Criteria for universality}

We have discussed  entanglement-based criteria for
non-encoded universality in Sec. \ref{sect_ED_criteria} and
\ref{sect_I_eff}. Here we discuss to which extent these
criteria can be applied---either directly or in an adapted
form---in the context of {\em encoded universality}
\footnote{We remark that this terminology must not be
confused with the notion of encoded universality used in
Ref. \cite{DiVi00}.}.

\subsubsection{Criteria at the level of encoded systems}

The first observation is that one  can use exactly the same
criteria for universality when considering encoded
universality, if one applies them at the level of encoded
systems. This is expressed in the following
\begin{thm}
Consider a fixed (consistent family of) encoding(s), which
also implies a partition of the system. Any
entanglement-based criterion for universality formulated
for non-encoded systems using LOCC leads to a criterion for
the encoded system by considering the corresponding
entanglement measures where locality is defined with
respect to the partition of the system, i.e., {\bf \em
L}OCC.
\end{thm}
The  proof of this theorem is straightforward, as one can
treat any logical qubit as a two--level system if {\bf
L}OCC are allowed.

\begin{rem} {\it Entanglement with respect to partitions.---}
For  a {\em fixed} encoding, one considers local operations
with respect to this encoding, i.e. {\bf L}OCC.
Entanglement is defined relative to the partitions induced
by the encoding. For instance, one says that $E_{\rm
Bell}(|\bm \phi\rangle) = 1$ if an encoded entangled pair
$(|{\bm 0}\rangle|{\bm 0}\rangle+ |{\bm 1}\rangle|{\bm
1}\rangle)/\sqrt{2}$ can be created between some pair of
encoded systems. In case of entanglement width measures,
they are now defined with respect to the fixed partition.
That is, entanglement between blocks of qubits is
considered, where each block consists of several physical
qubits specified by the encoding. Entanglement within each
block is irrelevant and absorbed into the encoding. Notice
that in this way one still obtains a set of powerful
criteria to assess whether a given state is encoded
universal with respect to a fixed encoding. In particular,
{\em any} encoded universal resource still needs to have
unbounded entanglement width (defined via {\bf L}OCC).
\hfill $\diamond$
 \end{rem}

\begin{rem} {\it Simultaneous treatment of different encodings.---}
Notice that  above criterion automatically provides a test
for all encodings that lead to the same partition of the
system, i.e. the same notion of {\bf L}OCC, independent of
the choice of the two orthogonal states constituting the
logical qubit. However, it would still be desirable to
obtain a more general statement about non-universality with
respect to {\em all} possible encodings. Although many
encodings are treated simultaneously, one still needs to
check for all possible partitions of the system and
determine the corresponding measures, which is impractical.
\hfill $\diamond$
\end{rem}

In the following, we utilize entanglement-based criteria at
the level of physical systems to infer statements about
encoded universality. This turns out to be possible,
although only under certain circumstances and for certain
entanglement measures.

\subsubsection{Schmidt measure and geometric measure}

In this section, we will show that every encoded universal
resource must have an unbounded Schmidt measure and
geometric measure, i.e., $E_s(\Psi)=\infty = E_g(\Psi)$ for
every encoded universal resource $\Psi$.

We start by considering the  geometric measure. Let
$|\psi\rangle$ be a state defined on a set of qubits
$S:=\{1, \dots, n\}$, and let ${\cal P}:=\{A_1, \dots,
A_l\}$ be any partition of the set $S$ into $l$ subsets
$A_i$. One can regard $|\psi\rangle$ as an $l$-party state,
where every subset $A_i$ is associated to a party. At this
coarse-grained level one can also define the geometric
measure $E_g^{\cal P}$ in a natural way, where now locality
is defined with respect to these $l$ parties. An important
property is now that $E_g^{\cal P}(|\psi\rangle)\leq
E_g(|\psi\rangle)$ for any partition ${\cal P}$, i.e., the
geometric measure is {\em non-increasing under coarsening
of the partition} \cite{Ei01}.

This property will be crucial for proving  that
$E_g(\Psi)=\infty$ for every encoded universal resource
$\Psi$. Let $\{|{\bm 0}\rangle, |{\bm 1}\rangle\}$ be an
$m$-qubit encoding with respect to which $\Psi$ is encoded
universal, and let $|\phi\rangle= \sum c_{i_1\dots
i_n}|i_1\dots i_n\rangle$ be an arbitrary $n$-qubit state.
By the encoded universality of $\Psi$, there exists a state
$|\psi_i\rangle\in\Psi$ and an encoded version
$|\bm\phi\rangle$ of $|\phi\rangle$ such that
$|\psi_i\rangle \geq_{\mbox{\tiny LOCC}} |\bm \phi\rangle$.
As $E_g$ is a type II entanglement monotone,  this implies
that $E_g(|\psi_i\rangle) \geq E_g( |\bm \phi\rangle)$. Let
$S:=\{1, \dots, m\cdot n\}$ be the set of qubits on which
$|\bm\phi\rangle$ is defined, and ${\cal P}$ be the
partition of $S$ which is naturally associated with the
encoded state. Using now the property that $E_g$ is
non-increasing under coarsening, one finds that $E_g( |\bm
\phi\rangle)\geq E_g^{\cal P}( |\bm \phi\rangle)$. Now, let
$U$ be an $m$-qubit unitary operation such that \be U |{\bm
0}\rangle = |0\rangle^{\otimes m}\mbox{ and } U  |{\bm
1}\rangle= |1\rangle|0\rangle^{\otimes (m-1)}, \ee implying
that $U^{\otimes n} |\bm \phi\rangle$ is equal to \be  \sum
c_{i_1\dots i_n} \left\{|i_1\rangle |0\rangle^{\otimes
(m-1)}\right\}\dots \left\{|i_n\rangle |0\rangle^{\otimes
(m-1)}\right\}.\ee This equation should be read in the
sense that the state $|\phi\rangle$ is distributed over $n$
single-qubit parties, where every party  supplemented with
an uncorrelated ancilla of $m-1$ qubits in the state
$|0\rangle^{\otimes (m-1)}$. As such ancillas cannot change
the value of any type I entanglement monotone \cite{Vi00},
one finds that $E_g^{\cal P}( U^{\otimes n}|\bm
\phi\rangle) = E_g(|\phi\rangle)$. Moreover, as $U^{\otimes
n}$ is a local operation with respect to the partition
${\cal P}$, one has  $E_g^{\cal P}( U^{\otimes n}|\bm
\phi\rangle)= E_g^{\cal P}( |\bm \phi\rangle)$. This proves
that $E_g(|\psi_i\rangle) \geq E_g( |\phi\rangle)$. As this
result holds for arbitrary $|\phi\rangle$, we find  that
$E_g(\Psi)=\infty$, as desired.

Note that in the proof of the above result we have only used that (i) $E_g$ is a type II entanglement monotone, (ii) $E_g$ is a type I entanglement monotone, and (iii) $E_g$ is non-increasing under coarsening. Therefore, the same arguments hold for the Schmidt measure $E_s$, and we can immediately formulate the following general result.

\begin{thm}\label{thm_encoded_nogo}
Let $E$ be a measure which is a type I and type II
entanglement monotone and which is non-increasing under
coarsening. Then $E(\Psi) = E^*$ for every encoded
universal resource $\Psi$. In  particular, $E_s(\Psi) =
E_g(\Psi)=\infty$ for every encoded universal resource.
\end{thm}

Note that the above result is general in  that it is
formulated independent of any specific encoding. It allows
to rule out e.g. the W states as resources which are
certainly not encoded universal (see also section
\ref{sect_I_gm}).

\subsubsection{Localizable entanglement}

We now turn to the criteria based on the  ability of
creating Bell pairs between qubits in the system. Recall
that we have seen in Sec. \ref{sect_I_le} that for any
universal resource it must be possible to create a Bell
state between some pair of qubits in the system  (recall
also that we consider fixed output particles, i.e., the
same for all branches of the protocol). The results of Ref.
\cite{Gr06} show that this criterion cannot be suitable to
assess universality of encoded resources (if one applies it
at the level of physical qubits). Although Ref. \cite{Gr06}
is only concerned with CC-universality, one can lift their
results to the more general level of encoded
CQ-universality, as we will see in this section.

The failure of the above measure is already  evident when
considering the following simple example based on the
results of \cite{Gr06}. Suppose one has the encoding \be
\label{Wencoding} |{\bm 0}\rangle = |0\rangle ^{\otimes m},
\hspace{1cm} |{\bm 1}\rangle = |W_m\rangle, \ee where
$|W_m\rangle$ is a W state of $m$ qubits, for some fixed
$m$.  A state \be |\phi\rangle  = 1/\sqrt{2} (|{\bm
0}\rangle|{\bm 0}\rangle+ |{\bm 1}\rangle|{\bm 1}\rangle)
\ee is evidently maximally entangled at the encoded level,
although at the level of the physical qubits only a certain
(arbitrary small) amount of (entropic) entanglement is
present. This can be seen by considering the reduced
density operator of a single individual particle $k$, which
is given by \be \rho_k= \frac{2m-1}{2m} |0\rangle\langle 0|
+  \frac{1}{2m} |1\rangle \langle 1|. \ee Clearly, the
entropy of entanglement can go to zero as $m \to \infty$.
The entanglement of assistance (and therefore also the
localizable entanglement) between any pair of qubits is
upper bounded by $S(\rho_k)$ \cite{Fo06}, and hence also
goes to zero for $m \to \infty$. However, maximal
entanglement between {\em blocks of qubits}, i.e., at the
encoded level, is naturally present.

As it was proven in Ref. \cite{Gr06} that the 2D  cluster
states encoded with respect to the encoding specified in
Eq. \ref{Wencoding} form a CC-universal resource (in fact,
we will prove below that this resource is also encoded
CQ-universal), one can conclude that measures related to
the possibility of creating Bell pairs are \emph{not}
suitable to study encoded universality. In particular, such
a resource is {\rm not} not CQ-universal, even though it is
encoded CQ-universal.

\subsubsection{Entanglement width}

We now turn to the entanglement width measures. We will use the result of Ref. \cite{Va06a} to establish a criterion for encoded universality (at the level of physical qubits). The criterion will be stated independent of any encoding, as is the case with theorem \ref{thm_encoded_nogo}.

As discussed earlier, in Ref. \cite{Va06a} it was  shown
that LOCC protocols on all (families of) states with at
most logarithmically growing $\chi$-width can be simulated
efficiently on a classical computer. This implies that, for
these resource states, measurement-based quantum
computation on encoded qubits can be simulated efficiently
on a classical computer for {\em any} choice of encoding.
This result does, strictly speaking, not exclude the
possibility that states with logarithmically bounded
$\chi$-width constitute a universal encoded resource, as it
has to date not been proven that an efficient classical
simulation of (CC-)universal quantum computation is
impossible. However, under the assumption that one will not
be able to efficiently simulate quantum computation, one
can conclude the following from the  theorem above.

\begin{thm}
Let $\Psi$ be a family of states such that the
$\chi$-width grows at most logarithmically with the system
size on the set $\Psi$. Then, under the assumption that
(CC) quantum computation cannot be simulated efficiently by
classical computers, the resource $\Psi$ cannot be encoded
universal.
\end{thm}

\begin{rem} {\it Direct use of entanglement width.---} So far we have not
 been able to directly prove (or disprove)  that every encoded universal
 resource should have an unbounded entropic entanglement width and/or Schmidt-rank
 width. This is mainly due to the fact that these measures do not seem to exhibit a
 'non-increasing under coarsening' property. In particular, it seems
 that these measures can (significantly) increase due to the use of encodings. Therefore,
 we leave this issue as an interesting open problem. \hfill $\diamond$
\end{rem}

\subsection{Encoded universal resources}

In this section, we prove a general result on encoded
universality up to logical local operations, which
immediately provides a large number of encoded universal
resources.

\begin{thm}\label{encodedEUR}
Let $\Psi$ be a family of states that is a universal
resource under local projective measurements, up to local
unitary operations. Then, for any encoding $\{|\bm
0\rangle, |\bm 1\rangle\}$, the encoded version of the
family, denoted by $\bm \Psi$, is an encoded universal
resource under LOCC up to logical local unitary operations.
\end{thm}
The proof of this theorem is a straightforward generalization of a particular example of an encoded universal resource presented in \cite{Gr06}. We use a result of Walgate et al. \cite{Wa00}, stating that any two orthogonal multi-qubit quantum states can be distinguished by LOCC deterministically. As a consequence, for any choice of encoding $\{|\bm 0\rangle = |\psi_0\rangle, |\bm 1\rangle =|\psi_1\rangle\}$, any two-outcome projective measurement within the logical subspace can be performed by means of LOCC, i.e., by operating on the physical qubits independently. To see this, consider a two-outcome projective measurement specified by the projectors onto the two orthogonal states $\{|{\phi}_0\rangle, |{\phi}_1\rangle\}$. Let $|{\bm \phi}_0\rangle, |{\bm \phi}_1\rangle$ be encoded versions of these states (which are therefore also orthogonal). Any encoded version $|\bm \psi\rangle$ of a state $|\psi\rangle$ can be written as
\be
|\bm \psi\rangle = |{\bm \phi}_0\rangle|{\bm \chi}_0\rangle + |{\bm \phi}_1\rangle|{\bm \chi}_1\rangle,
\ee
where $|{\bm \chi}_0\rangle,|{\bm \chi}_1\rangle$ are non-normalized, possible non-orthogonal, encoded states of the remaining system. Since one can  deterministically distinguish $|{\bm \phi}_0\rangle$ and $|{\bm \phi}_1\rangle$ by performing local measurements only on the particles corresponding to these states, the state of the remaining (i.e., unmeasured) particles is given by $|{\bm \chi}_0\rangle/\sqrt{p_0}$ or $|{\bm \chi}_1\rangle/\sqrt{p_1}$, depending on the measurement outcome. The success probabilities for the two branches are given by $p_0 :=  |\langle{\bm \chi}_0|{\bm \chi}_0\rangle|^2$ and $p_1 :=  |\langle{\bm \chi}_1|{\bm \chi}_1\rangle|^2$.
Note that the unmeasured system is in the same state as if the logical projective measurement had been performed on the logical system, and the corresponding probabilities to obtain the outcomes are also the same. Note also that the measured system is not in an eigenstate of the measured logical observable. However, this is irrelevant as we are no longer interested in states of measured systems but rather in state of the remaining ones. Consider now a sequence of projective measurements that would produce a certain output state $|\phi_{\rm out}\rangle$ up to local (i.e., single qubit) operations from a given resource state. If we perform the encoded versions of this sequence of measurements on the encoded state, i.e., measurements on the logical subspace, we obtain as output state an encoded version of the state $|\bm \phi_{\rm out}\rangle$ up to logical single system operations, as desired.

Notice  that the result is general in the sense that it
covers {\em all} possible encodings, and for any encoding
it is sufficient to perform only {\em local} operations on
the physical qubits. As an immediate consequence, we obtain
the following.

\begin{thm}
For any encoding, the family of encoded 2D-cluster states $\{|\bm C_{d \times d}\rangle\}$ is an encoded universal resource up to single system logical Pauli operations under LOCC. Any family of encoded 2D-cluster states is also $CC$-universal under LOCC.
\end{thm}

The proof is again straightforward. We apply the same sequence of single qubit measurements as in the one--way model for MQC on the level of encoded, i.e. logical system. This can be done by using only LOCC as shown above. If the initial state was an encoded 2D--cluster state, the resulting state $|\bm \phi_{\rm out}\rangle$ is an encoded version of the desired output state $|\phi_{\rm out}\rangle$ up to logical single system Pauli operations, similar as in the one--way model.

\begin{rem} {\it Performing logical Pauli operations by LOCC.---}
We remark that it is necessary to consider (logical  single
system) correction operations at the end of the process, as
is, e.g., required in the one-way model. Otherwise it may
often not be possible to generate the desired output state
deterministically. Notice that it is not always possible to
perform the logical Pauli operations, i.e. the final basis
change, by means of LOCC. This is also the reason why the
above theorems can only be stated up to logical Pauli
operations. However, if we are interested in classical
output data only (CC-universality), these logical Pauli
operations correspond to a known basis change and do not
play a role. They simply correspond to an adjustment of the
final measurement basis, where again the same techniques to
perform projective measurements on logical systems by means
of LOCC can be used.

An example of an encoding where logical Pauli operations
cannot be performed locally is given by Eq.
(\ref{Wencoding}). In this case, $\bm \sigma_x$ would need
to map a product state onto an entangled state, which is
clearly impossible by LOCC. For certain encodings, logical
Pauli operations {\em can} however be implemented by LOCC.
For such encodings, the resulting resource is fully
CQ-encoded universal. An example for such an encoding is
provided by \be \label{GHZ encoding} |\bm 0\rangle
=|0\rangle^{\otimes m}; \hspace{1cm} |\bm
1\rangle=|1\rangle^{\otimes m}. \ee Here we have   ${\bm
\sigma_x}= \sigma_x^{\otimes m}$, ${\bm \sigma_z}=
\sigma_z^{(1)}$, ${\bm \sigma_y}=
\sigma_y^{(1)}\sigma_x^{\otimes (m-1)}$. It might be
interesting to look for other encodings where this is the
case. Natural candidates are encodings where the logical
basis states are obtained from each other by LOCC, i.e.
have the same type of entanglement.\hfill $\diamond$
\end{rem}

\subsection{Discussion of encoded universal resources in the literature}

Some examples of encoded universal resources, often as
proposals for physical implementation of one-way
computation, have been discussed previously in the
literature. In this subsection we focus on  encodings that
are naturally covered in our framework. We do not intend to
review all potential implementations here; a brief summary
is available in Sec. 5 of Ref.~\cite{He06}.

In Ref.~\cite{Gr06}, Gross and Eisert describe a specific
example of an encoded universal resource using the encoding
(\ref{Wencoding}). The focus of their investigations lies
on CC-universality, and only the classical information of
measurement outcomes performed at the end of a one-way
computation is considered. These authors use a different
terminology to describe measurement-based quantum
computation, and methods from many-body physics (such as matrix product states and their generalizations to 2D systems) are used and extended. The abovementioned example the authors
propose fits into the general framework described in the
present paper. In particular, as one finds (see Theorem
\ref{encodedEUR}) for general encodings, the resource is
also CQ-universal up to logical Pauli operations.
Additional examples of universal resources based on
weighted graph states are put forward in Ref. \cite{Gr06}.
The example presented in Fig. 1(a) of Ref. \cite{Gr06} fits
into the framework we consider as well, and this resource
fulfills the entanglement-based criteria for
CQ-universality we presented above, even without
considering encodings. The relation, if there is any, between the other examples given in Ref.~\cite{Gr06} and the current investigation, is at present unclear, and understanding whether there is a connection between the two approaches would be an interesting topic for further research. 

The authors of Ref.
\cite{Gr06} stress that their resources have radically
different entanglement properties than the (unencoded)
cluster state. This is, however, only true when considering
``local'' entanglement features with respect to physical
qubits. ``Global'' entanglement properties (such as
entanglement width, geometric measure, Schmidt measure) as
well as entanglement width respect to the proper encoding
(i.e. {\bf L}OCC) seem to be in fact the {\em same} as of a
2D cluster state, as we have illustrated in this section.

Another example of an encoded universal resources is given
in Ref.~\cite{Ba06}, where Bartlett and Rudolph show how to
obtain encoded graph states corresponding to square and
hexagonal lattices as the (approximate) ground state of a
nearest-neighbor two-body interaction Hamiltonian. The
encoding they use is given by Eq.~(\ref{GHZ encoding}), and
they show that MQC can proceed on the encoded cluster
states under LOCC (i.e., by physical single-qubit
measurements). The focus of Ref. \cite{Ba06} lies on the
realization of an encoded universal resource state by means
of two-body Hamiltonians with a constant gap between the
ground state and excited states, regardless of the system
size. Note that, in contrast, the generation of a
non-encoded graph states as the exact ground state of a
two-body Hamiltonian is impossible (see
Ref.~\cite{Va06GS}).

Encodings into quantum error correction codes are naturally
utilized for fault-tolerant computation, and the simulation
of fault-tolerant circuits (i.e., circuits on the encoded
logical qubits by quantum codes) with 2D cluster states has
been discussed in Ref.~\cite{FTQC}. Recently, in
Ref.~\cite{Fu06}, fault-tolerant quantum computation with
an encoded 2D cluster state is discussed. The scheme is
based on a high degree of verification in terms of
error-detection (and post-selection) using a
Calderbank-Shor-Steane code. Hence, the logical qubits are
themselves stabilizer states, so that this specific
encoding allows implementation of logical ${\bm \sigma}_z$
measurements by (transversal) single qubit measurements,
and logical Clifford gates by (transversal) single--qubit
gates. The latter property makes arbitrary logical Clifford
measurements possible. The missing gate for universality, a
$\pi/8$ gate, is provided by preparing a logical qubit in
the corresponding site in a specific state, namely
$\cos(\pi/8)|{\bf 0}\rangle + \sin(\pi/8) |{\bf 1}\rangle$,
together with (transversal) single-qubit measurements. We
remark that our approach (via Theorem 17) in principle
provides an alternative way to perform the required logical
single-system measurements and hence universal MQC,
although it is not clear whether fault-tolerance is
guaranteed. In Ref. \cite{Ra06}, entirely new schemes for
fault-tolerant quantum computation using topologically
protected quantum information (surface codes) have been
proposed. We remark that such types of global encodings may
not be treated in a straightforward manner by the methods
developed in this article.

\section{Summary}\label{sect_summary}

In this paper, we have investigated which states constitute
universal resources for measurement-based computation
(MQC), and what the role of entanglement is in this issue.

\subsection{Considerations regarding universality in QC and MQC}

Sections \ref{General} and \ref{sect_universal_resource}
were dedicated to investigating what is exactly meant by
``universal quantum computation'', and, more specifically,
``universal \emph{one-way} quantum computation''. We argued
that a distinction needs to be made between several types
of universality---called CC-, QC-, CQ- and
QQ-universality---depending on whether the input and output
of a quantum computer are allowed to be either classical
(C) or quantum (Q). In  the current investigation of
universality in MQC, we focused on \emph{CQ-universality},
due to  the following reasons.
\begin{itemize}
\item[(i)] Among the above four types of universality,
CQ-universality is the most general type any (one-way) MQC
model can have. In  particular, any CQ-universal resource
is also CC-universal. The reason that, e.g.,
QQ-universality is not possible, is essentially due to the
fact that in one-way models, where resource states are
processed with \emph{local} operations only,  it is not
possible to have quantum states as inputs; only classical
inputs are possible.

\item[(ii)] Although the most general property of
QQ-universality  is not achievable in any MQC model where
resource states can be processed by LOCC only, we have
shown that a small modification of the scheme---a
supplementary ``input coupler'' in the form of a sequence
of phase gates or Bell measurements, which is used only at
the beginning of an algorithm to read in quantum
inputs---suffices to make any CQ-universal MQC model also
QQ-universal.

\item[(iii)] Furthermore, as we have shown here, a
systematic treatment of CQ-universality---i.e., formulation
of criteria for CQ-universality as well as construction of
new universal resource states---is possible. Such an
investigation reveals a central role of entanglement.
\end{itemize}
Nevertheless, it is clear that other types of entanglement
also deserve a detailed investigation. In  particular, if
one is interested in quantum computers as devices which
only provide solutions to classical problems (such as,
e.g., prime factoring), one is evidently interested in
CC-universality in MQC, rather than CQ- or QQ-universality.
However, systematic treatments of CC-universality, and the
role of entanglement in such issues, seem much less clear.

\subsection{Criteria for universality}

In sections \ref{sect_obs}, \ref{sect_ED_criteria} and
\ref{sect_I_eff}, we  have developed a framework to
investigate (CQ$-$)universality in MQC. We have analyzed
which entanglement features any (efficient) universal
resource must have. We considered type II monotones $E$,
which are non-increasing under deterministic LOCC
conversion, and found that every universal resource must
reach the supremum of any such measure $E$. Thus, universal
resources are maximally entangled if entanglement is
quantified by such measures $E$. This insight leads to a
class of criteria for universality, as any resource which
does not exhibit such a maximal entanglement, cannot be
universal. We have illustrated this strategy by considering
the entanglement width measures, measures related to
localizable entanglement, the geometric measure and the
Schmidt measure, and have obtained the following criteria
for universality (theorems \ref{thm_ED_ewd},
\ref{thm_ED_le}, \ref{thm_ED_gm}, \ref{thm_ED_sm}).

\

\noindent {\bf Criteria for universality.}  {\it For
every universal resource,  the following measures must be
unbounded: \begin{itemize}\item[(i)] entropic entanglement
width $E_{\mbox{\scriptsize wd}}$; \item[(ii)] Schmidt-rank
width $\chi_{\mbox{\scriptsize wd}}$; \item[(iii)]  maximal
size ${\cal N}_{\mbox{\scriptsize{LE}}}$ of a set of qubits with pairwise unit localizable
entanglement;\item[(iv)]
geometric measure $E_g$, and \item[(v)] Schmidt measure
$E_s$. \end{itemize}}

Subsequently, several examples of non-universal  resources
have been given. We name the 1D cluster states, W states,
GHZ states and ground states of non-critical 1D spin
systems, as most important examples.

When efficiency is additionally required, the above
criteria can be strengthened, as one finds that the scaling
behavior (with respect to the number of qubits) of all
measures $E$ on efficient universal resources must be
sufficiently strong:

\

\noindent {\bf Criteria for efficient universality.}
{\it For any efficient universal resource,  the measures
$E_{\mbox{\scriptsize wd}}$, $\chi_{\mbox{\scriptsize
wd}}$, ${\cal N}_{\mbox{\scriptsize{LE}}}$, $E_g$ and $E_s$ must
scale faster than logarithmically with the number of
qubits.}

\

We noted that these results agree with recent results
stating that MQC can in fact be classically simulated
efficiently on all resources where the Schmidt-rank width
grows at most logarithmically with the system size.

It is important to make the following
comment. When assessing the universality or
non-universality of a given resource, the choice of
suitable entanglement measures $E$ may strongly depend on
the resource which is studied. In other words, every
measure has a regime of (families of) states in the Hilbert
space, where it is most powerful. This is most distinctly
illustrated when assessing the universality of graph state
resources. One finds that e.g. the measures related to
localizable entanglement are \emph{not} useful to
investigate universality of graph states---since all (fully
entangled) graph states have maximal localizable
entanglement between every pair of qubits in the system
\cite{He06}, and therefore localizable entanglement does
not distinguish between, e.g, the 1D and the 2D cluster
states. Similarly, the geometric measure and the Schmidt
measure fail to detect the 1D cluster states as being
non-universal, as these measures grow unboundedly on these
states. This reflects the fact that all graph states are
highly entangled in several ways. On the other hand, the
entanglement width measures do turn out to be very powerful
to classify graph states in the context of universality,
and are e.g. able to identify 1D cluster states (and
several other examples) as non-universal. Hence, we can
conclude that an important aspect of this study is to map
out, for a given measure $E$, the regime where this measure
is useful to detect non-universal resources. Conversely, it
is important, for a given resource or class of resources,
to identify those measures $E$ which are useful to asses
their universality.

\subsection{Examples of universal resources}

In section \ref{go}, we have presented examples of
efficient universal resources (see also Ref. \cite{Va06}),
namely the graph state resources associated to the 2D
hexagonal, triangular and Kagome lattices. We have provided
explicit LOCC protocols which efficiently transform these
states to the 2D cluster states.

\subsection{Encoded universality}

In section \ref{encoded}, we considered a more general
notion of CQ-universality, called \emph{encoded
CQ-universality}, in MQC. A resource is called encoded
universal if it is possible to generate arbitrary encoded
states by performing LOCC on the resource. There are
several reasons for considering encoded universality:

\begin{itemize}
\item[(i)] For many applications (e.g., situations where
qudit  states are represented by encoded qubit states, as
well as  in fault tolerant schemes), it is sufficient---and
even preferred---that output states are generated in
encoded form.

\item[(ii)] Encoded universality is a weaker notion of
CQ-universality than the one considered above---i.e., every
CQ-universal resource is also encoded CQ-universal. In
particular, by considering encoded universality in MQC one
comes closer to the notion of CC-universality, such that
insight in the former might lead to insights in the latter.

\item[(iii)] A systematic investigation of encoded CQ-universality using considerations regarding entanglement, leading both to  criteria for encoded universality and to new constructions of universal resources, is possible.
\end{itemize}

We subsequently studied which criteria have to be fulfilled by any encoded universal resource. Most importantly, we found that the Schmidt measure and geometric measure of entanglement must be unbounded on every encoded universal resource.

Finally, we provided a class of constructions  of encoded
universal resources, by showing that every CQ-universal
resource, which is subsequently encoded, is also an encoded
universal resource (up to logical Pauli operations). In
particular, such resources are also fully CC-universal.

\section{Outlook and open problems}\label{sect_outlook}

In this paper, we have treated ``exact and deterministic
universality'',  i.e., we have considered the case where
the desired output states are created with unit probability
(i.e. deterministically) and with unit accuracy (i.e.
exactly). In the present study, we have not considered
the---from a practical perspective---relevant case of {\em
approximate} and {\em non--deterministic} generation of the
output states. In a forthcoming paper \cite{Mo07}, we will
extend the definition of universality to take these weaker
requirements into account, and investigate to which extent
the entanglement-based criteria established for exact
deterministic universality can be transferred to this more
general case. While some measures (most notably local
measures such as localizable entanglement) can no longer be
used, many of the present results carry over to this more
general case.
In Ref. \cite{Mo07}, we will also provide examples of
approximate quasi-deterministic universal resources that
are not exact deterministic universal, including states
different from graph states.

We remark that several questions in the present study
remain to date unanswered, most notably the issue whether
the entanglement width measures can be used directly to
assess encoded universality. Furthermore, additional
insights are needed regarding the scaling of entanglement
in efficient encoded universal resources. In  a broader
scope, it would be interesting to understand whether
criteria for efficient CC-universality differ from the
criteria for efficient (encoded) CQ-universality, i.e.,
whether there exist efficient CC-universal resources that
are not (encoded) CQ-universal.

We are convinced that insights gained in the investigation
of the universality of resources for measurement-based
quantum computation allow us to deepen our understanding of
the nature and the potential of quantum computation.

\section*{Acknowledgments}
For enlightening discussions, we thank R. Werner, R. Jozsa, R. Raussendorf, G. Vidal, S. Aaronson, S.-I. Oum, F. Verstraete, J. Eisert, D. Gross, D. E. Browne, T. Short,  S. Flammia, N. Schuch, E. Kashefi, D. Markham, V. Vedral, J. Anders, and C. Mora.  We thank an anonymous referee for pointing out remark 12.
This work was supported by the Austrian Science Foundation (FWF),
the European Union (OLAQUI, SCALA, QICS), the Austrian Academy of Sciences
(\"OAW) through project APART (W.D.), and JSPS (A.M.).

%

\end{document}